\documentclass[preprint]{elsarticle}
\usepackage[margin=3cm]{geometry}
\usepackage{lineno,hyperref}
\modulolinenumbers[5]
\usepackage{amssymb}
\usepackage{subfigure}

\usepackage{amsmath} 
\usepackage{color} 
\usepackage{mathtools} 

\usepackage[]{algorithm2e} 

\newcommand{\mb}[1] {\mathbf{#1}}                           
\newcommand{\il}[0] {\int\limits}                           
\newcommand{\suml}[0] {\sum\limits}                           

\newcommand{\realS}[0] {``real space" sum}
\newcommand{\kspaceS}[0] {``$\mb{k}$-space" sum}
\newcommand{\realP}[0] {``real space" part}
\newcommand{\kspaceP}[0] {``$\mb{k}$-space" part}
\newcommand{\kspace}[0] {``$\mb{k}$-space"}
\newcommand{\rspace}[0] {``real space"}

\newcommand{\G}[0] {\mathcal{G}}
\newcommand{\GR}[0] {\G^R}
\newcommand{\GF}[0] {\widehat{\G}^F}
\newcommand{\Gself}[0] {\G_{\textrm{self}}}

\newcommand{\Hd}[0] {\mathcal{H}}
\newcommand{\HdR}[0] {\Hd^R}
\newcommand{\HdF}[0] {\widehat{\Hd}^F}
\newcommand{\Hdself}[0] {\Hd_{\textrm{self}}}

\newcommand{\Kop}[0] {\mathcal{K}}

\newcommand{\R}[0] {\mathcal{R}}
\newcommand{\rect}[1] {\text{rect}\left({#1}\right)}

\newcommand{\Kx}[2] {K_{#1}\left( {#2} \right)}
\newcommand{\Kxy}[3] {K_{#1}\left( {#2},{#3} \right)}
\newcommand{\Ex}[2] {E_{#1}\left( {#2} \right)}

\newcommand{\GFR}[0] {\widehat{\G}^{F,\R}}
\newcommand{\HdFR}[0] {\widehat{\Hd}^{F,\R}}

\newcommand{\uf}[0] {u^{f}}
\newcommand{\up}[0] {u^{p}}

\newcommand{\ufF}[0] {u^{f,F}}

\newcommand{\Jx}[2] {J_{#1}\left( {#2} \right)}

\newcommand{\D}[0] {\mathcal{D}}

\makeatletter
\def\ps@pprintTitle{%
 \let\@oddhead\@empty
 \let\@evenhead\@empty
 \def\@oddfoot{}%
 \let\@evenfoot\@oddfoot}
\makeatother

\begin{document}

\begin{frontmatter}

\title{Spectrally accurate Ewald summation for the Yukawa potential in two dimensions}
\author[add1]{Sara P\aa lsson \corref{cor1}}
\ead{sarapal@kth.se}
\author[add1]{Anna-Karin Tornberg}

\cortext[cor1]{Corresponding author}
\address[add1]{KTH Mathematics, Linn\'e Flow Centre, 100 44 Stockholm Sweden}

\begin{abstract}
An Ewald decomposition of the two-dimensional Yukawa potential and its derivative is presented for both the periodic and the free-space case. These modified Bessel functions of the second kind of zeroth and first degrees are used e.g. when solving the modified Helmholtz equation using a boundary integral method. The spectral Ewald method is used to compute arising sums at $\mathcal{O}(N\log N)$ cost for $N$ source and target points. To facilitate parameter selection, truncation-error estimates are developed for both the real-space sum and the Fourier-space sum, and are shown to estimate the errors well.
\end{abstract}

\begin{keyword}
Ewald summation \sep Bessel functions \sep Yukawa potential \sep periodicity \sep fast summation
\end{keyword}

\end{frontmatter}



\section{Introduction}
This paper concerns the fast and accurate computation of discrete sums containing the Yukawa potential and its first derivative. The Yukawa potential is the free-space Green's function for the modified Helmholtz equation and is also known as the screened Coulomb potential. It is also the zeroth order modified Bessel function of the second kind and it is often denoted $\Kx{0}{r}$. The modified Helmholtz equation arises in a number of applications, from the solution of the forced isotropic heat equation \cite{fryklund2019} and in the linearisation of the Poisson-Boltzmann equation \cite{cheng2006}, among others. For the quasi-two-dimensional case, i.e. the two-dimensional case using the three-dimensional definition of the Yukawa potential, an Ewald split for the Yukawa potential has been derived in \cite{mazars2007}.

A boundary integral formulation can be used to solve the modified Helmholtz equation \cite{cheng2006,fryklund2019,Kropinski2011a}. The double-layer potential in such a formulation contains the normal derivative of the Yukawa potential and hence the first-order modified Bessel function of the second kind, $\Kx{1}{r}$. Using a suitable discretisation scheme, at the core it remains to evaluate discrete sums with $\Kx{0}{r}$ and $\Kx{1}{r}$. These sums need to be evaluated for all target points where a solution is required. In the case of a periodic setting, special techniques are needed to facilitate their efficient computation. In a free-space setting, the sums can be computed directly at cost $\mathcal{O}(N^2)$ for $N$ sources and targets. As problem sizes grow, the evaluation becomes costly.
There are several ways to speed up the computations of both the periodic and the free-space sums: for non-periodic boundary conditions, often the Fast Multipole Method (FMM) \cite{Greengard1997} is  used and was presented in \cite{cheng2006,Kropinski2011a} for the modified Helmholtz equation.
Another approach to compute these sums is to use FFT-based methods \cite{darden1993, Lindbo2011a}, which is especially suitable for periodic problems.
One such method is the spectral Ewald method which has been implemented for Laplace's equation in 3D \cite{AfKlinteberg2014,Lindbo2011a} as well as for the Stokes equations in 3D with periodic \cite{Lindbo2010} and non-periodic \cite{AfKlinteberg2017a} boundary conditions. For 2D, the spectral Ewald method has been used to compute solutions to the Stokes equations in a periodic setting \cite{Palsson2019}. Using an Ewald decomposition of the sum into a \realP~which converges rapidly in real space and a \kspaceP~which converges rapidly in Fourier space, the spectral Ewald method allows the two sums to be computed in $\mathcal{O}(N\log N)$ time.
The approach is valid for both the periodic and the free-space case, with only minor differences \cite{AfKlinteberg2017a}.

In this paper, an Ewald decomposition for $\Kx{0}{r}$ and $\Kx{1}{r}$ is derived and the spectral Ewald method is presented for both the periodic and the free-space problems. Truncation-error estimates are developed to facilitate parameter selection and the computational complexity is shown to be $\mathcal{O}(N\log N)$. It is shown how further speed ups can be achieved for cases when the target points are located on a uniform grid, which is the case e.g. when solving the modified Helmholtz equation in \cite{fryklund2019}.

The paper is organised as follows: below is a short summary of a boundary integral formulation for the modified Helmholtz equation as a motivation to this work. In \S\ref{sec:ewald} the Ewald decomposition is derived for both $\Kx{0}{r}$ and $\Kx{1}{r}$ in both the periodic and the free-space settings. The spectral Ewald method is described in \S\ref{sec:se} for both cases. In \S\ref{sec:est}, truncation-error estimates are derived for both the \realS~and the \kspaceS.

\subsection{Motivation}
The sums containing the Yukawa potential and its derivative, $\Kx{0}{r}$ and $\Kx{1}{r}$, arise e.g. from a boundary integral formulation of the modified Helmholtz equation. The equation with Dirichlet boundary conditions takes the form
\begin{align}
\begin{split}
  \alpha^2 u(\mb{x}) - \Delta u(\mb{x}) &= 0, \; \mb{x}\in\Omega, \\
  u(\mb{x}) &= g(\mb{x}), \; \mb{x}\in\Gamma,
\end{split}
\label{eq:modhelm}
\end{align}
on a domain $\Omega$ with boundary $\Gamma$ and $u$ unknown in $\Omega$. The free-space Green's function for the operator $\alpha^2 - \Delta$ is defined as
\begin{align}
  G(\mb{x},\mb{y}) = \alpha^2 K_0\left( \alpha|\mb{y}-\mb{x}| \right),
  \label{eq:Gorig}
\end{align}
where $\Kx{0}{r}$ is the zeroth order modified Bessel function of the second kind. A solution to \eqref{eq:modhelm} is given by the double layer formulation
\begin{align*}
  u(\mb{x}) = \dfrac{\alpha^2}{4\pi^2} \il_\Gamma M(\mb{x},\mb{y})\mu(\mb{y})ds_y, \; \forall \mb{x}\in\Omega,
\end{align*}
where
\begin{align*}
  M(\mb{x},\mb{y}) = -\dfrac{\partial}{\partial \nu_y} K_0\left( \alpha|\mb{y}-\mb{x}| \right) = \alpha K_1\left( \alpha|\mb{y}-\mb{x}| \right) \dfrac{\mb{y}-\mb{x}}{|\mb{y}-\mb{x}|}\cdot \mb{n}_y.
\end{align*}
The outward normal, $\mb{n}_y$ at point $\mb{y}$ can here be regarded as known. If the boundary condition in \eqref{eq:modhelm} is changed from a Dirichlet condition to a Neumann condition, the single layer formulation with $K_0$ is needed to evaluate the solution in the domain. In short, the sums to evaluate in the free-space setting take the form
\begin{align}
   \uf_{\G}(\mb{x}) &\coloneqq \suml_{n=1}^N  \G (\mb{x},\mb{y}_n) f_{\G}(\mb{y}_n), \label{eq:sumG} \\
  \uf_{\Hd}(\mb{x}) &\coloneqq \suml_{n=1}^N \Hd(\mb{x},\mb{y}_n) \cdot \mb{f}_{\Hd}(\mb{y}_n) \label{eq:sumH},
\end{align}
where
\begin{align}
  \G(\mb{x},\mb{y}) \coloneqq& \,K_0\left( \alpha|\mb{y}-\mb{x}| \right), \label{eq:G} \\
\Hd(\mb{x},\mb{y})\coloneqq& \,K_1\left(\alpha|\mb{y}-\mb{x}|\right) \dfrac{\mb{y}-\mb{x}}{|\mb{y}-\mb{x}|},
\label{eq:H}
\end{align}
and $N$ is the number of source points $\mb{y}_n$.
The functions $f_{\G}(\mb{y}_n)\coloneqq \frac{\alpha^2}{4\pi^2}\mu(\mb{y}_n)w_n ds_n$ and $\mb{f}_{\Hd}(\mb{y}_n) \coloneqq \frac{\alpha}{2\pi}\mu(\mb{y}_n)w_n n_{y_n}ds_n$ come from the discretisation of the single and double layer formulations respectively, with $w_k$ being the quadrature weight corresponding to $\mb{y}_k$, and $ds_k$ a line segment.

For some applications, the solution $u(\mb{x})$ needs to be evaluated on a uniform grid in $\Omega$. When solving the heat equation \cite{fryklund2019}, this is needed to facilitate the application of 2D FFTs. For other applications, the target points $\mb{x}$ are distributed along a boundary. The method proposed in this paper must be able to handle both cases efficiently. Complexity, efficiency and depndency on certain parameters is discussed in \S\ref{sec:res}.


\section{Ewald decomposition}
\label{sec:ewald}

\begin{figure}[h!]
  \centering
  \includegraphics[width=0.3\textwidth]{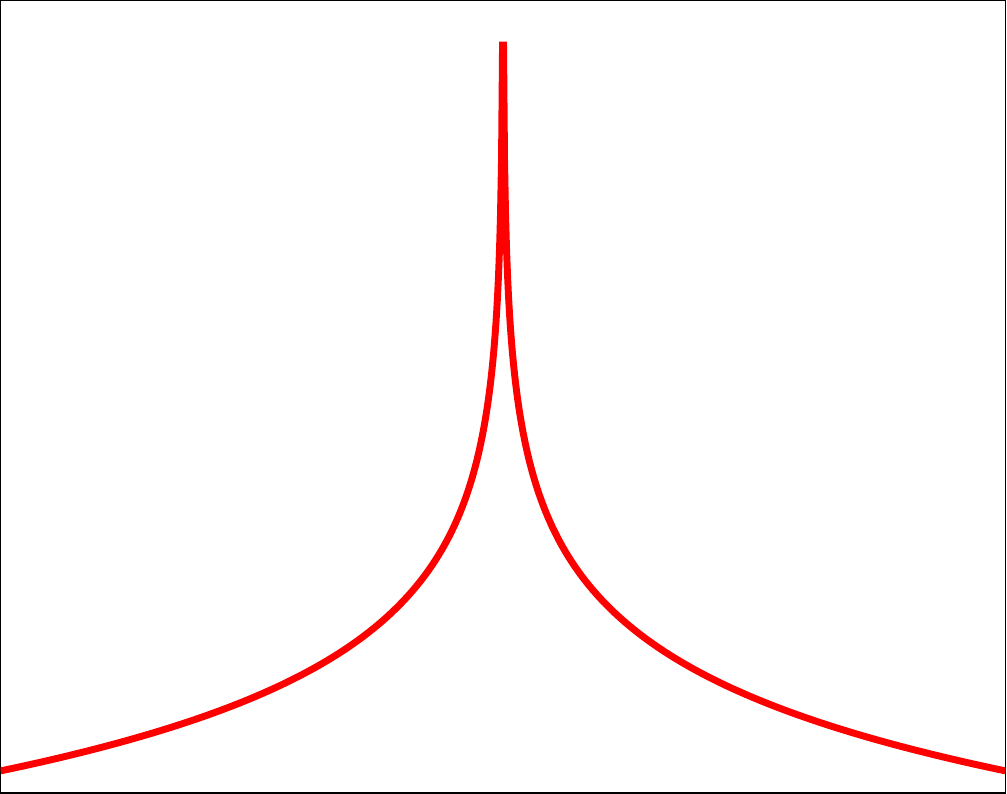}%
  \hspace{1cm}
  \includegraphics[width=0.3\textwidth]{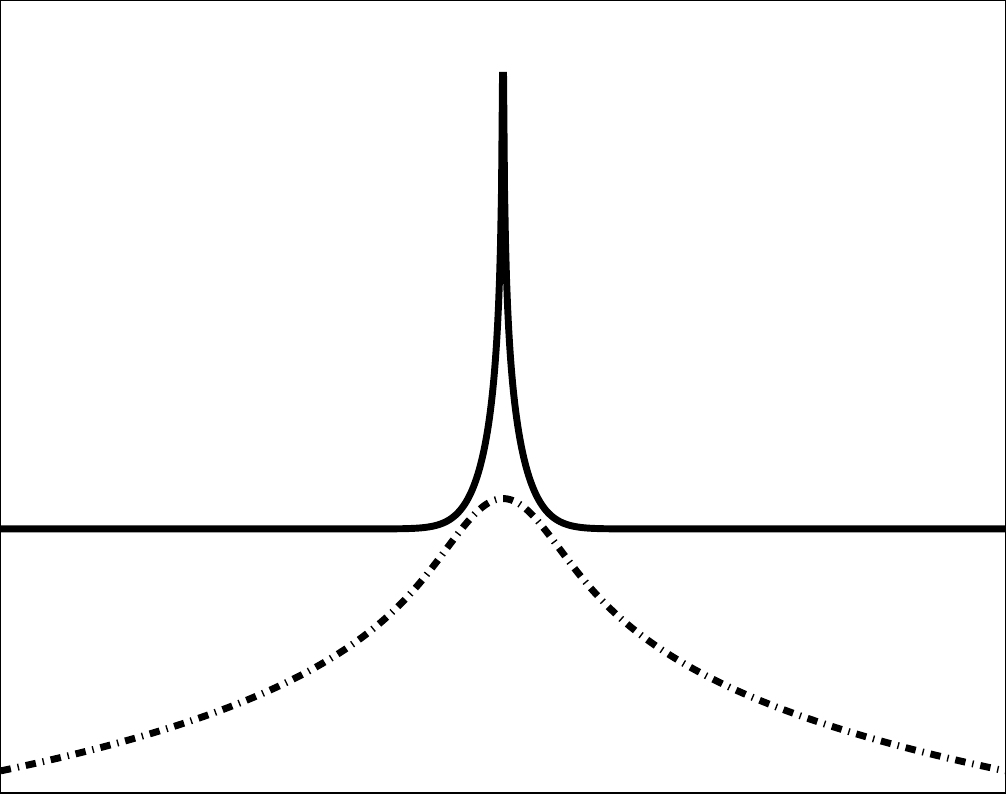}%
  \caption{Left: the harmonic Green's function $\mathcal{L}(r)$. Right: the decomposition of $\mathcal{L}$ into $\mathcal{L}^R$ (solid black line) and $\mathcal{L}^F$ (dashed black line) for one specific choice of decomposition parameter $\xi$. }
  \label{fig:ewald_L}
\end{figure}

This section describes the decomposition of $\G(\mb{r})\coloneqq \G(\mb{x},\mb{y})$ and $\Hd(\mb{r})\coloneqq \Hd(\mb{x},\mb{y})$ for $\mb{r}=\mb{y}-\mb{x}$, to compute the sums in \eqref{eq:sumG} and \eqref{eq:sumH} fast and with spectral accuracy.
The decomposition is derived in a similar way to the derivation for the Stokeslet ant stresslet in \cite{Palsson2019}.
The starting point is the classic approach by \citeauthor{Ewald1921} \cite{Ewald1921} to decompose a slowly converging sum into two: one rapidly converging sum in real space and one sum which converges rapidly in Fourier space. The sums are called the \realS~and \kspaceS~respectively. To demonstrate the idea of such a split, the harmonic Green's function $\mathcal{L}(r)=-\log(\vert r \vert)/2\pi$ will be studied briefly. As shown in Figure~\ref{fig:ewald_L} (left), the behaviour of $\mathcal{L}$ is long-ranging.
It is split using the classical Ewald screening function, $\gamma(r,\xi) = \xi^2e^{-\xi^2 r^2}/\pi^2$. The split can be computed as
\begin{align*}
  \mathcal{L}^R(r,\xi) =  \dfrac{E_1(\xi^2r^2)}{4\pi}, \quad \mathcal{L}^F(r,\xi) = -\dfrac{\log(\vert r \vert)}{2\pi}-\dfrac{E_1(\xi^2r^2)}{4\pi},
\end{align*}
where $\mathcal{L}^R$ and $\mathcal{L}^F$ are denoted the \realP~and \kspaceP~respectively. The parameter $\xi$ governs the balance between the \realP~and the \kspaceP. The split is computed using techniques derived in \cite{Palsson2019}. The decomposed functions are shown in Figure~\ref{fig:ewald_L} (right), where the short-range behaviour of $\mathcal{L}^R$ is clear and the \kspaceP~is smooth and slow-varying, which translates into a short-ranging behaviour in Fourier space.

\subsection{Periodic setting}
In a periodic setting with a periodic domain $\Omega$ of size $L_1\times L_2$, $u_{\G}(\mb{x})$ from \eqref{eq:sumG} is extended as
\begin{align}
  \up_{\G}(\mb{x}) = \suml_{\mb{p}\in\mathbb{Z}^2} \suml_{n=1}^N \G(\mb{x}-\tau(\mb{p})-\mb{y}_n)f_{\G}(\mb{y_n}),
  \label{eq:sumG_per}
\end{align}
where $\tau(\mb{p})=(p_1 L_1, p_2 L_2)^T$ for $\mb{p}=(p_1, p_2)^T$, $p_1,p_2\in\mathbb{Z}$. Here, $\mb{y}_n$, $n=1,\hdots,N$ are the set of source points with strengths $f_{\G}(\mb{y}_n)$. The aim of an Ewald decomposition is to split the expression in \eqref{eq:sumG_per} into two: a \realS~and \kspaceS~such that
\begin{align}
  \up_{\G}(\mb{x}) = \suml_{\mb{p}\in\mathbb{Z}^2}^{*} \suml_{n=1}^N \GR\left( \mb{x}-\tau(\mb{p})-\mb{y}_n,\xi \right)f_{\G}(\mb{y}_n) +
  \dfrac{1}{V}\suml_{\mb{k}} \GF\left(\mb{k},\xi\right) \suml_{n=1}^N f_{\G}(\mb{y}_n)e^{-i\mb{k}\cdot(\mb{x}-\mb{y}_n)},
  \label{eq:sumG_per_split}
\end{align}
where the asterisk denotes that the term $\mb{x}-\tau(\mb{p})-\mb{y}_n=0$ has been excluded from the sum and $V=L_1L_2$. Since that term should also be excluded from the \kspaceS, in the case of a target point $\mb{x}=\mb{y}_n$ for some $n$, the limit
\begin{align*}
  \lim_{|\mb{r}|\rightarrow 0} \left(\GR(\mb{r},\xi))-G(\mb{r})\right)f(\mb{y}_n),
\end{align*}
should be added to \eqref{eq:sumG_per_split}. A similar decomposition is sought for the periodic extension of \eqref{eq:sumH}. The term $\mb{k}=0$ in the \kspaceS~sets the constant of the periodic solution. It is here chosen to use the term $\GF(0,\xi)$ without further modifications, as it is well-defined for both $\G$ and $\Hd$. This choice makes the split independent on the parameter $\xi$.

\subsubsection{Decomposition of $\G(\mb{r})$}
The Green's function to split is defined in \eqref{eq:G}, for $r = |\mb{r}|$ and $\mb{r}=\mb{x}-\mb{y}$. It is the fundamental solution to the equation $\left( \alpha^2 \mb{I} - \Delta\right)\G = 2\pi\delta$ and its Fourier transform is defined as
\begin{align}
  \widehat{\G}(k) = \dfrac{2\pi}{\alpha^2 + k^2},
  \label{eq:Ghat}
\end{align}
for $k=|\mb{k}|$ and $\mb{k}=(k_1, k_2)$ and $k_j = 2\pi\kappa_j/L$ for $\kappa_j\in\mathbb{Z}$ for both $j=1$ and $2$.

The computation of $\up_{\G}(\mb{x})$  in \eqref{eq:sumG_per} can be seen as the solution to the problem
\begin{align*}
  (\alpha^2\mb{I}-\Delta)\phi(\mb{x}) = 2\pi\suml_{n=1}^N \sigma^n(\mb{x}),
\end{align*}
with
\begin{align*}
  \sigma^n(\mb{x}) \coloneqq \sum_{\mb{p}\in\mathbb{Z}^2}f_{\G}(\mb{y}_n)\delta(\mb{x}- \mb{y}_n- \tau(\mb{p})).
\end{align*}
Here, $\phi(\mb{x}) \coloneqq \up_{\G}(\mb{x})$ for ease of notation.

To decompose $\phi(\mb{x})$, the Ewald screening function \cite{Ewald1921} is used, with a slight modification to account for the $\alpha$-term, i.e.
\begin{align}
  \gamma_\alpha (\mb{x},\xi) = \dfrac{\xi^2}{\pi}e^{-\alpha^2/4\xi^2}e^{-\xi^2 \vert\mb{x}\vert^2} \Leftrightarrow \widehat{\gamma_\alpha}(\mb{k},\xi) = e^{-(\alpha^2+\vert\mb{k}\vert^2)/4\xi^2},
  \label{eq:screening}
\end{align}
The source term $\sigma^n$ is split accordingly;
\begin{align}
  \sigma^n(\mb{x}) = \underbrace{\sigma^n(\mb{x})-\left(\sigma^n*\gamma\right)(\mb{x})}_{\eqqcolon \sigma^{n,R}(\mb{x})} + \underbrace{\left(\sigma^n*\gamma\right)(\mb{x})}_{\eqqcolon \sigma^{n,F}(\mb{x})}.
  \label{eq:sigmadef}
\end{align}
By the linearity of the operator $\alpha^2\mb{I}-\Delta$, $\phi(\mb{x})$ can be decomposed as $\phi(\mb{x})=\sum_{n=1}^N \phi^{n,R}+\phi^{n,F}$ where
\begin{align*}
  (\alpha^2\mb{I}-\Delta)\phi^{n,F}(\mb{x}) &= 2\pi \sigma^{n,F}(\mb{x}), \\
  (\alpha^2\mb{I}-\Delta)\phi^{n,R}(\mb{x}) &= 2\pi \sigma^{n,R}(\mb{x}).
\end{align*}
To compute $\phi^{n,F}$, the problem is considered in the frequency domain with $\phi^{n,F}=\sum_{\mb{k}} \widehat{\phi}^{n,F} e^{i\mb{k}\cdot\mb{x}}$. It can be written
\begin{align*}
  (\alpha^2\mb{I}-\Delta)\phi^{n,F}(\mb{x}) = \suml_{\mb{k}} (\alpha^2+k^2)\widehat{\phi}^{n,F}(\mb{k})e^{i\mb{k}\cdot\mb{x}},
\end{align*}
where $k=\vert\mb{k}\vert$.
By the Poission summation formula, it holds that
\begin{align*}
  \sigma^{n,F} = \dfrac{f_{\G}(\mb{y}_n)}{V}\suml_{\mb{k}}\widehat{\gamma}(\mb{k},\xi)e^{i\mb{k}\cdot(\mb{x}-\mb{y}_n)}.
\end{align*}
By orthogonality, equating the two expressions gives
\begin{align*}
  \phi^{n,F}(\mb{x})= \dfrac{f_{\G}(\mb{y}_n)}{V}\suml_{\mb{k}} \dfrac{2\pi}{\alpha^2+k^2}\widehat{\gamma}(\mb{k},\xi)e^{i\mb{k}\cdot(\mb{x}-\mb{y}_n)}.
\end{align*}
Comparing $\phi^{n,F}$ with the expressions in \eqref{eq:sumG_per_split}, it is clear that
\begin{align}
  \GF(-\mb{k},\xi) = \dfrac{2\pi}{\alpha^2+k^2}\widehat{\gamma}(\mb{k},\xi) = \dfrac{2\pi}{\alpha^2+k^2}e^{-(\alpha^2+k^2)/4\xi^2}.
  \label{eq:GF_per}
\end{align}
The real space part, $\phi^{n,R}$, is defined as
\begin{align*}
  \phi^{n,R}(\mb{x}) &= \sigma^n*(\delta-\gamma)(\mb{x}) \\
  &= \il_{\mathbb{R}^2} f_{\G}(\mb{y}_n)\G(\mb{y}-\mb{x})\suml_{\mb{p}\in\mathbb{Z}^2}\left(\delta(\mb{y}-\mb{y}_n-\tau(\mb{p}))-\gamma(\mb{y}-\mb{y}_n-\tau(\mb{p}))\right)d\mb{y} \\
  &\eqqcolon \suml_{\mb{p}\in\mathbb{Z}^2}f_{\G}(\mb{y}_n)\GR(\mb{x}-\mb{y}_n-\tau(\mb{p},\xi).
\end{align*}
To compute $\GR$ directly is somewhat tricky. Instead of directly computing the convolution integral, for which a closed formulation has not been found, it is first expressed in Fourier space as
\begin{align*}
  \widehat{\GR}(\mb{k},\xi) = \widehat{\G}(k) - \widehat{\G}(k)\widehat{\gamma_\alpha}(k,\xi) = \dfrac{2\pi}{\alpha^2 + k^2}\left(1-e^{-(\alpha^2+k^2)/4\xi^2}\right).
\end{align*}
The inverse Fourier transform can be expressed as
\begin{align*}
  \GR(\mb{r},\xi) = \dfrac{1}{4\pi^2} \il_{\mathbb{R}^2} \widehat{\GR}(k,\xi)e^{i\mb{k}\cdot\mb{r}}d\mb{k} = \dfrac{1}{2\pi} \il_{\kappa=0}^\infty \il_{\theta=0}^{2\pi} \dfrac{1-e^{-(\alpha^2+\kappa^2)/4\xi^2}}{\alpha^2+\kappa^2}e^{i\kappa r \cos(\theta-\beta)}\kappa d\kappa d\theta,
\end{align*}
where in the last step polar coordinates are introduced such that $(k_1,k_2) = \kappa(\cos(\theta),\sin(\theta))$ and $\mb{r}=r(\cos(\beta),\sin(\beta))$. Integrating over $\theta$ gives
\begin{align*}
  \GR(\mb{r},\xi) = \underbrace{\il_0^\infty \dfrac{\kappa}{\alpha^2+\kappa^2}J_0(\kappa r)d\kappa}_{\eqqcolon I_1} - \underbrace{\il_0^\infty \dfrac{\kappa}{\alpha^2+\kappa^2}J_0(\kappa r)e^{-(\alpha^2+\kappa^2)/4\xi^2}d\kappa}_{\eqqcolon I_2},
\end{align*}
where the first integral is evaluated to $I_1 = \Kx{0}{\alpha r}$. The second integral is more difficult to evaluate, and a similar trick to that in \cite{Tornberg2016} is used. First, let $\lambda=1/4\xi^2$ and compute
\begin{align*}
  \dfrac{\partial I_2}{\partial \lambda} = -e^{-\alpha^2\lambda}\il_0^\infty (\alpha^2+\kappa^2)\dfrac{\kappa}{\alpha^2+\kappa^2}J_0(\kappa r)e^{-\kappa^2\lambda}d\kappa = -e^{-\alpha^2\lambda}\dfrac{e^{-r^4/4\lambda}}{2\lambda}.
\end{align*}
Integrating $I_2$ with respect to $\lambda$ gives
\begin{align*}
  I_2 = \il_0^\lambda -e^{-\alpha^2\lambda}\dfrac{e^{-r^4/4\lambda}}{2\lambda}d\lambda = \dfrac{1}{2}\il_\lambda^\infty \dfrac{e^{-\alpha^2\eta}e^{-r^2/4\eta}}{\eta}d\eta =
   \dfrac{1}{2}\il_1^\infty \dfrac{e^{-\alpha^2t/4\xi^2}e^{-r^2\xi^2/t}}{t}dt=\dfrac{1}{2}
   \Kxy{0}{\dfrac{\alpha^2}{4\xi^2}}{r^2\xi^2},
\end{align*}
where the integration limits can be changed since $\lim_{\lambda\rightarrow\infty}  -e^{-\alpha^2\lambda}\dfrac{e^{-r^4/4\lambda}}{2\lambda}=0$.
Setting $z=\alpha^2/4\xi^2$ and $\omega=r^2\xi^2$, $\Kxy{0}{z}{\omega}$ is the incomplete modified Bessel function of the second kind of zeroth order. The definition for an integer order, $\nu$, is \cite{Harris2008}
\begin{align}
  \Kxy{\nu}{z}{\omega} = \il_1^\infty \dfrac{e^{-zt-\omega/t}}{t^{\nu+1}}dt.
  \label{eq:Kxy_def}
\end{align}
Together, the two integrals give the \realP~of $\Kx{0}{\alpha r}$ as
\begin{align}
  \GR(\mb{r},\xi) = \Kx{0}{\alpha r} - \dfrac{1}{2}\Kxy{0}{\dfrac{\alpha^2}{4\xi^2}}{r^2\xi^2} = \dfrac{1}{2}\Kxy{0}{r^2\xi^2}{\dfrac{\alpha^2}{4\xi^2}}.
  \label{eq:GR_per}
\end{align}
Here the relation
\begin{align}
  \Kxy{0}{z}{\omega} = 2\Kx{0}{2\sqrt{z\omega}}-\Kxy{0}{\omega}{z},
  \label{eq:Kxy_switch0}
\end{align}
has been utilised \cite{Harris2008}. The self-interaction term is computed by
\begin{align}
  \Gself(\mb{r},\xi) = \lim_{|\mb{r}|\rightarrow 0} \GR(\mb{r},\xi)-\G(\mb{r}) = \lim_{|\mb{r}|\rightarrow 0} -\dfrac{1}{2\pi}\Kxy{0}{\dfrac{\alpha^2}{4\xi^2}}{r^2\xi^2} = -\dfrac{1}{2\pi}\Ex{1}{\dfrac{\alpha^2}{4\xi^2}},
  \label{eq:Gself_per}
\end{align}
where the relation in \eqref{eq:Kxy_switch0} again has been used together with  $\Kxy{\nu}{z}{0}=\Ex{\nu+1}{z}$ \cite{Harris2008}.

\subsubsection{Decomposition of $\Hd(\mb{r})$}
To obtain a decomposition of $\Hd(\mb{r})$ as defined in \eqref{eq:H}, first the relation between $\G$ and $\Hd$ needs to be explored. Given that $\frac{\partial K_0}{\partial r}(r) = -\Kx{1}{r}$ it holds that
\begin{align*}
  -\nabla \Kx{0}{\alpha r} = \alpha \dfrac{\mb{r}}{r}\Kx{1}{\alpha r},
\end{align*}
i.e. the operator $\Kop \coloneqq -\frac{1}{\alpha}\nabla$ connects $\G$ and $\Hd$ as
\begin{align}
  \Hd(\mb{r}) = \Kop G(\mb{r}).
\label{eq:GHrel}
\end{align}
To obtain the Ewald decomposition of $\Hd$ it is thus enough to apply the operator $\Kop$ on the \realP~and \kspaceP~of $\G$ respectively \cite{AfKlinteberg2017a}. The periodic expression corresponding to \eqref{eq:sumH} is defined as
\begin{align}
  \up_{\Hd}(\mb{x}) = \suml_{\mb{p}\in\mathbb{Z}^2} \suml_{n=1}^N \Hd(\mb{x}-\tau(\mb{p})-\mb{y}_n)\cdot\mb{f}_{\Hd}(\mb{y_n}).
  \label{eq:sumH_per}
\end{align}
The Ewald decomposition of \eqref{eq:sumH_per} thus reads
\begin{align}
  \up_{\Hd}(\mb{x}) = \suml_{\mb{p}\in\mathbb{Z}^2}^{*} \suml_{n=1}^N \HdR\left( \mb{x}-\tau(\mb{p})-\mb{y}_n,\xi \right)f_{\G}(\mb{y}_n) +
  \dfrac{1}{V}\suml_{\mb{k}} \HdF\left(\mb{k},\xi\right) \cdot \suml_{n=1}^N \mb{f}_{\Hd}(\mb{y}_n)e^{-i\mb{k}\cdot(\mb{x}-\mb{y}_n)},
  \label{eq:sumHd_per_split}
\end{align}
where the first sum corresponds to the \realS~and the second to the \kspaceS. The \realP~is obtained as
\begin{align}
  \HdR(\mb{r},\xi) = \Kop\GR(\mb{r},\xi) = \dfrac{\xi^2}{\alpha}\mb{r}\Kxy{-1}{r^2\xi^2}{\dfrac{\alpha^2}{4\xi^2}},
  \label{eq:HR_per}
\end{align}
where $K_{-1}$ is defined in \eqref{eq:Kxy_def}. Here, the following relation \cite{Harris2008} has been used,
\begin{align}
    \frac{\partial}{\partial z} \Kxy{0}{z}{\omega} &= -\Kxy{-1}{z}{\omega},
  \label{eq:Kxy_diff}
\end{align}
To obtain the \kspaceP, first the pre-factor that is produced when applying $\Kop$ to $e^{-i\mb{k}\cdot\mb{r}}$ needs to be computed:
\begin{align*}
  \Kop e^{-i\mb{k}\cdot\mb{r}} = \dfrac{i}{\alpha}\mb{k}e^{-i\mb{k}\cdot\mb{r}} \; \Rightarrow \; \widehat{\Kop} = \dfrac{i}{\alpha}\mb{k}.
\end{align*}
Thus, it is given that
\begin{align}
  \HdF(-\mb{k},\xi) = \widehat{\Kop}\GF(-\mb{k},\xi) = \dfrac{2\pi i}{\alpha} \dfrac{\mb{k}}{\alpha^2+k^2}e^{-(\alpha^2+k^2)/4\xi^2}.
  \label{eq:HF_per}
\end{align}
The self-interaction term is computed through the limit
\begin{align}
  \Hdself = \lim_{|\mb{r}|\rightarrow 0} \HdR(\mb{r},\xi)-\Hd(\mb{r}) = \lim_{|\mb{r}|\rightarrow 0} \left(\dfrac{-\xi^2 \mb{r}}{\alpha} \Kxy{1}{\dfrac{\alpha^2}{4\xi^2}}{r\xi^2}\right) = 0.
  \label{eq:Hself_per}
\end{align}
To compute the limit the relations
\begin{align*}
  \Kxy{1}{z}{\omega} = 2\sqrt{z/\omega}\Kx{1}{2\sqrt{z\omega}}-\Kxy{-1}{\omega}{z},
\end{align*}
and $\Kxy{1}{z}{0} = E_2(z)$ \cite{harris2009} are used.

\subsection{Free-space setting}
When considering the free-space case a similar decomposition can be introduced. In this case the expression in the \realS~remains unchanged, however the sum over the periodic replicas is removed. The discrete Fourier sum in the \kspaceS~is replaced by an inverse Fourier transform, i.e. \eqref{eq:sumG_per_split} becomes instead
\begin{align}
  \uf_{\G}(\mb{x}) = \suml_{n=1}^N \GR(|\mb{x}-\mb{y}_n|,\xi)f_{\G}(\mb{y}_n) +
   \underbrace{\dfrac{1}{4\pi^2}\il_{\mathbb{R}^2} \GF(\mb{k},\xi)\suml_{n=1}^N f_{\G}(\mb{y}_n)e^{i\mb{k}\cdot(\mb{y}_n-\mb{x})}d\mb{k}}_{\eqqcolon \ufF_{\G}},
  \label{eq:sumG_fs_split}
\end{align}
and $\uf_{\Hd}$ is similarly defined. To devise a numerical method, the integral in \eqref{eq:sumG_fs_split} needs to be discretised. As $\GF$ contains the factor $\frac{1}{\alpha^2 + k^2}$, care needs to be taken for small $\alpha$, where the integral is nearly singular for $k=0$. In these cases, a simple discretisation with the trapezoidal rule will not yield accurate results. To rectify this, truncations of $\G$ and $\Hd$ can be introduced, which are described in \S\ref{sec:sefs}.

The following tables summarise the Ewald decompositions: Table~\ref{tab:ewaldK0} lists the \realP~and \kspaceP~ for $\Kx{0}{\alpha r}$ and Table~\ref{tab:ewaldK1} for $\Kx{1}{\alpha r}$.

\begin{table}[h!]
  \centering
  \begin{tabular}{c | c }
    \hline
    \rspace & \kspace\\ \hline
    $\GR(r,\xi) = \dfrac{1}{2}\Kxy{0}{r^2\xi^2}{\dfrac{\alpha^2}{4\xi^2}}$ &
    $\GF(k,\xi) = \widehat{\G}(k) e^{-(\alpha^2+k^2)/4\xi^2}$
  \end{tabular}
  \caption{Summary of the obtained Ewald decompositions for $\Kx{0}{\alpha r}$ from \eqref{eq:GR_per} and  \eqref{eq:GF_per}, where $\widehat{\G}(k)$ is defined in \eqref{eq:Ghat}.}
  \label{tab:ewaldK0}
\end{table}

\begin{table}[h!]
  \centering
  \begin{tabular}{c | c }
    \hline
    \rspace & \kspace \\ \hline
    $\HdR(\mb{r},\xi) = \dfrac{\xi^2}{\alpha}\mb{r}\Kxy{-1}{r^2\xi^2}{\dfrac{\alpha^2}{4\xi^2}}$ &
    $\HdF(\mb{k},\xi) = \dfrac{-i\mb{k}}{\alpha}\widehat{\G}(k) e^{-(\alpha^2+k^2)/4\xi^2}$
    \end{tabular}
  \caption{Summary of the obtained Ewald decompositions for $\Kx{1}{\alpha r}$ from \eqref{eq:HR_per} and  \eqref{eq:HF_per}, where $\widehat{\G}(k)$ is defined in \eqref{eq:Ghat}.}
  \label{tab:ewaldK1}
\end{table}


\section{The Spectral Ewald method}
\label{sec:se}
The sums in \eqref{eq:sumG} and \eqref{eq:sumH} (and their periodic counterparts \eqref{eq:sumG_per} and \eqref{eq:sumH_per}) have been split into sums that converge rapidly:
the \realS~(\eqref{eq:GR_per} and \eqref{eq:HR_per}) in real space and the \kspaceS~(\eqref{eq:GF_per} and \eqref{eq:HF_per}) in Fourier space. However, without the application of fast summation methods, to compute these sums directly remains $\mathcal{O}(N^2)$ in complexity.
As a remedy, the spectral Ewald method is applied \cite{AfKlinteberg2014,Lindbo2010,Lindbo2011a}. The purpose of the method is to speed up computations to make the \realS~$\mathcal{O}(N)$ in cost and the \kspaceS~$\mathcal{O}(N\log N)$. How this is achieved has been described thoroughly in the references above for the three-dimensional case. As the method translates easily to two dimensions, here only a quick overview is presented. The balance between the \realS~and the \kspaceS~is governed by the splitting parameter $\xi$. The periodic domain is denoted $\Omega$ and is here assumed to be a square of size $L\times L$ for ease of notation.

\subsection{Fast real space summation}
The \realP~can for the periodic case be written as
\begin{align*}
  u^{R,P}(\mb{x}_t) &\coloneqq \suml_{\mb{p}\in\mathbb{Z}^2}^{*}\suml_{n=1}^N A^R(\mb{x}_t-\tau(\mb{p})-\mb{y}_n,\xi)f(\mb{y}_n), \; t=1,\hdots,N,
\end{align*}
where $A^R$ can be either $\GR$ or $\HdR$, a general form for \eqref{eq:Grealsum} and \eqref{eq:Hrealsum}. In the free-space case, the sum over $\mb{p}$ is removed. Evaluating these sums have $\mathcal{O}(N^2)$ complexity for $N$ target points $\mb{x}_t$.

As a first step, a cut-off radius $r_c$ introduced. Then, only the near neighbours are considered for each evaluation point $\mb{x}_t$, i.e. for $\text{NL}_t \coloneqq \left\{ (\mb{x}_s,\mb{p}):|\mb{x}_t-\mb{x}_s-\tau(\mb{p})|<r_c\right\}$,
\begin{align*}
  u^{R,P}(\mb{x}_t) \approx\suml_{(\mb{y},\mb{p})\in\text{NL}_t} A^R(\mb{x}_t-\mb{y}-\tau(\mb{p}),\xi)f(\mb{y}).
\end{align*}
With an Ewald decomposition as derived in \S\ref{sec:ewald} this sum is rapidly converging, but it remains $\mathcal{O}(N^2)$ to evaluate.

The complexity can be reduced to $\mathcal{O}(N)$ by modifying $r_c$ with increasing $N$ such that $\vert\text{NL}_t\vert$ remains constant. To create the neighbour lists $\text{NL}_t$ also has $\mathcal{O}(N)$ cost, which is achieved by first creating a linked cell list as in \cite{frenkel1997} for molecular dynamics. To keep the truncation errors at a desired tolerance when modifying $r_c$, the splitting parameter $\xi$ needs to be modified. In practice, changing $\xi$ will shift work from the \realS~to the \kspaceS. In order to select $r_c$ and $\xi$ to meet certain truncation errors, truncation-error estimates are derived in \S\ref{sec:est}.

\subsection{Fast Fourier space summation}
\label{sec:se_Fper}
The \kspaceS,
\begin{align}
  u^{F,P}(\mb{x}_t) &\coloneqq \dfrac{1}{L^2}\suml_{\mb{k}} \widehat{A}^F(\mb{k},\xi)\suml_{n=1}^N f(\mb{y}_n)e^{-i\mb{k}\cdot(\mb{x}_t-\mb{y}_n)}, \; t=1,\hdots N,
  \label{eq:se_persum}
\end{align}
where $\widehat{A}^F$ is either $\GF$ or $\HdF$ can be rewritten as
\begin{align*}
    u^{F,P}(\mb{x}_t) = \dfrac{1}{L^2}\suml_{\mb{k}} \widehat{A}^F(\mb{k},\xi)e^{-i\mb{k}\cdot\mb{x}_t}\suml_{n=1}^N f(\mb{y}_n)e^{i\mb{k}\cdot\mb{y}_n}, \; t=1,\hdots N.
\end{align*}
The aim is to compute the sum $u^{F,P}$ efficiently. It is beneficial to use FFTs on a uniform grid for the computations. However, the non-uniform source points $\mb{y}_n$ first need to be spread to the grid. The uniform grid on $\Omega$ is defined to have $M^2$ points with grid size $h=L/M$. The spreading is achieved using a window function, $w(\mb{x},\xi,\eta)$ where $\eta>0$ is some scaling parameter. The Fourier transform of $w(\mb{x},\xi,\eta)$ is denoted $\widehat{w}(\mb{k},\xi,\eta) = \widehat{w}(|\mb{k}|,\xi,\eta)\eqqcolon \widehat{w}_k$.
The expression above can be written as
\begin{align}
    u^{F,P}(\mb{x}_t) = \dfrac{1}{L^2}\suml_{\mb{k}} \widehat{A}^F(\mb{k},\xi)e^{-i\mb{k}\cdot\mb{x}_t}\dfrac{1}{\widehat{w}_k^2}\suml_{n=1}^N f(\mb{y}_n)\widehat{w}_k^2e^{i\mb{k}\cdot\mb{y}_n}, \; t=1,\hdots N,
    \label{eq:se_ufp_w}
\end{align}
where a factor $\widehat{w}_k^2/\widehat{w}_k^2$ has been introduced.
Denoting
\begin{align*}
  \widehat{H}(\mb{k}) \coloneqq \suml_{n=1}^N f(\mb{y}_n)\widehat{w}_k e^{-i\mb{k}\cdot\mb{y}_n},
\end{align*}
gives
\begin{align*}
    u^{F,P}(\mb{x}_t) = \dfrac{1}{L^2}\suml_{\mb{k}} \widehat{A}^F(\mb{k},\xi)e^{-i\mb{k}\cdot\mb{x}_t}\dfrac{\widehat{w}_k}{\widehat{w}_k^2} \widehat{H}(-\mb{k}), \; t=1,\hdots N.
\end{align*}
The function $\widehat{H}(\mb{k})$ is the result of a convolution in real space where
\begin{align}
  H(\mb{x}) = \suml_{n=1}^N f(\mb{y}_n)w(\mb{x}-\mb{y}_n),
  \label{eq:se_H}
\end{align}
which is a smooth function in the domain. $H(\mb{x})$ is used to spread the function $f$ on the sources to the uniform grid. In the case when $f$ has more than one component, all are spread to the grid independently. Once $H(\mb{x})$ is computed, $\widehat{H}$ can be obtained using a 2D FFT. It can be scaled by computing
\begin{align}
  \widehat{\tilde{H}}(\mb{k}) \coloneqq \widehat{A}^F(-\mb{k},\xi)\dfrac{\widehat{H}(\mb{k})}{\widehat{w}_k^2},
  \label{eq:se_Htildek}
\end{align}
such that
\begin{align*}
    u^{F,P}(\mb{x}_t) = \dfrac{1}{L^2}\suml_{\mb{k}} \widehat{\tilde{H}}(-\mb{k}) \widehat{w}_k e^{-i\mb{k}\cdot\mb{x}_t}, \; t=1,\hdots N.
\end{align*}
For periodic functions, it holds that
\begin{align*}
  \suml_{\mb{k}} \widehat{h}(-\mb{k})\widehat{g}(\mb{k}) = \il_{\Omega}h(x)g(x)dx.
\end{align*}
Identifying $\widehat{h}\coloneqq\widehat{\tilde{H}}$ and $\widehat{g}\coloneqq\widehat{w}_ke^{-i\mb{k}\cdot\mb{x}_t}$, where $\widehat{g}$ is the result of a convolution, it is obtained that
\begin{align}
  u^{F,P}(\mb{x}_t) = \il_{\Omega} \tilde{H}(\mb{y})w(\mb{x}_t-\mb{y})_*d\mb{y},
  \label{eq:se_perint}
\end{align}
where $\Omega$ is the periodic reference domain. The asterisk denotes that  periodicity is implied in both directions.
Assuming the window function has compact support, the integral can then be evaluated using the trapezoidal rule as the integrand is smooth and periodic. Then, $\tilde{H}(\mb{y})$ is needed for $\mb{y}$ on the uniform grid and can be obtained using the inverse FFT. The window function is assumed to have support on $p$ points in each direction.

Given a set of sources $\mb{y}_n$, $n=1,\hdots,N$ and targets $\mb{x}_t$, $t=1,\hdots,N$, to compute $u^{F,P}(\mb{x}_t)$ in \eqref{eq:se_persum} for all $t$ consists of the following steps \cite{AfKlinteberg2014}:
\begin{enumerate}
  \item \emph{Spreading}: compute $H(\mb{x})$ on the uniform grid with $M^2$ points. This involves evaluating $\mathcal{O}(N)$ window functions on $p^2$ points, at cost $\mathcal{O}(p^2N)$.
  \item \emph{FFT}: compute $\widehat{H}(\mb{k})$ using the 2D FFT, at cost $\mathcal{O}(M^2\log M)$.
  \item \emph{Scaling}: compute the tensor-product $\widehat{\tilde{H}}(\mb{k})$ for all $k_1,k_2\in[-M/2,M/2-1]$, which is of $\mathcal{O}(M^2)$ complexity.
  \item \emph{IFFT}: compute $\tilde{H}(\mb{x})$ on the uniform grid using the 2D inverse FFT, at cost $\mathcal{O}(M^2\log M)$.
  \item \emph{Quadrature}: the integral in \eqref{eq:se_perint} needs to be evaluated for all target points to obtain $u^{F,P}(\mb{x}_t)$. Using the compactly supported window functions, this is of cost $\mathcal{O}(p^2N)$.
\end{enumerate}
One example of a window function is the truncated Gaussian, where $w(\mb{x},\xi,\eta) = w^1(x,\xi,\eta)w^1(y,\xi,\eta)$ for
\begin{align*}
  w^1(x,\xi,\eta)= \begin{cases} e^{-\eta |x|^2/\omega^2}, & ,|x|\leq \omega = \frac{ph}{2}, \\
  0, & \text{otherwise}, \end{cases}
\end{align*}
where $\eta$ is a scaling parameter. To further reduce the cost of the spreading and quadrature steps, in the case of Gaussians for window functions, Fast Gaussian Gridding \cite{Greengard2004} is applied. See \cite{Lindbo2011a} for notes on implementation. The truncated Gaussian is the window function used in this paper.

\subsection{Parameter selection}
To keep the evaluation of the real space sum $\mathcal{O}(N)$, the cut-off radius $r_c$ is set initially, given the number of sources and targets. A truncation-error estimate is used to choose a suitable $\xi$. Given $\xi$, an appropriate $k_\infty = M/2$ is computed from a \kspace~truncation-error estimate. The truncation-error estimates of both $\G$ and $\Hd$ are derived in \S\ref{sec:est} for both the \realP~and the \kspaceP. The support of the window functions, $p$, is chosen large enough to keep the approximation errors low.  The parameter $\eta$ balances the errors from truncation and approximation, i.e. how well the Gaussians are resolved on the grid. Here, it is chosen that $\eta = 0.95^2 \pi p/2$.

\subsection{Free-space case}
\label{sec:sefs}
Above, the spectral Ewald method for the periodic case is explained. For the free-space case, the expression instead has the form \eqref{eq:sumG_fs_split}. The aim is to compute the integral in $u_{\G}^{f,F}$ using the trapezoidal rule and accelerate the computations using FFTs. Following the approach of \cite{vico2016}, the Fourier transform of $\G$ is regularised by cutting off the interaction in physical space beyond the domain of interest, here denoted $\D$. The same approach has been used to derive free-space Ewald decompositions of the Stokeslet and stresslet in 3D \cite{AfKlinteberg2017a}.

Considering the equation
\begin{align*}
  (\alpha^2 \mb{I}-\Delta)\phi(\mb{x}) = 2\pi g(\mb{x}),
\end{align*}
where $g(\mb{x})$ has compact support on $\D$ and with the free-space boundary condition $\phi(\mb{x})\rightarrow 0$ as $\vert\mb{x}\vert\rightarrow \infty$, the solution can be written as
\begin{align}
  \phi(\mb{x}) = \il_{\mathbb{R}^2}\G(\vert \mb{x}-\mb{y}\vert)g(\mb{y})d\mb{y}.
  \label{eq:freespace_phi}
\end{align}
The Green's function $\G$ can be modified to
\begin{align}
  \G^{\R}(\mb{r}) = \G(\mb{r})\rect{\dfrac{|\mb{r}|}{2\R}},
  \label{eq:Gmod}
\end{align}
where
\begin{align*}
  \rect{x} = \begin{cases} 1, &\; |x|\leq 1/2, \\
  0, &\; |x|>1/2, \end{cases}
\end{align*}
and $\R$ is larger than the largest point-to-point distance in the domain $\mathcal{D}$. For $g$ compactly supported in $\mathcal{D}$, the solution within $\D$ is not changed by considering $\G^{\R}$ instead of $\G$. Following the approach by \cite{vico2016}, the Fourier transform of $\G^{\R}$ is obtained through
\begin{align}
  \widehat{\G^{\R}}(k) = 2\pi \il_0^\infty \Jx{0}{kr}\G^{\R}(r)r dr = \dfrac{2\pi}{\alpha^2+k^2}\left[ 1 + \alpha k \Jx{1}{k\R}\Kx{0}{\alpha\R} -\alpha\R\Jx{0}{k\R}\Kx{1}{\alpha\R} \right].
  \label{eq:GhatR}
\end{align}

The modified Green's function \eqref{eq:Gmod} replaces \eqref{eq:G} to compute the \kspaceP~in the free-space case. Together with the screening function,
\begin{align}
 \GFR(\mb{k},\xi) = \dfrac{2\pi}{\alpha^2+k^2}\left[ 1 + \alpha k \Jx{1}{k\R}\Kx{0}{\alpha\R} -\alpha\R\Jx{0}{k\R}\Kx{1}{\alpha\R} \right]e^{-(\alpha^2+k^2)/4\xi^2},
 \label{eq:GF_fs}
\end{align}
which replaces $\GF$ in \eqref{eq:sumG_fs_split} and gives an identical result in $\mathcal{D}$.
To resolve the integral with $\GFR$ is easier for small $\alpha$ than with the original $\GF$, since $\GFR$ has a well-defined limit for $k=0$ when $\alpha\rightarrow 0$. The behaviour of both $\GF$ and $\GFR$ is demonstrated in \S\ref{sec:results_modG}.
The \realP, $\GR$, and the self-interaction part, $\Gself$, remain unchanged.
Similarly, to obtain the \kspaceP~of $\Hd(\mb{r})$ in the free-space case, the operator $\widehat{\Kop}$ is applied on $\GFR(\mb{k},\xi)$ which gives
\begin{align}
 \HdFR(\mb{k},\xi) = \dfrac{2\pi i}{\alpha}\dfrac{\mb{k}}{\alpha^2+k^2}\left[ 1 + \alpha k \Jx{1}{k\R}\Kx{0}{\alpha\R} -\alpha\R\Jx{0}{k\R}\Kx{1}{\alpha\R} \right]e^{-(\alpha^2+k^2)/4\xi^2}.
 \label{eq:HF_fs}
\end{align}
With these expressions, similar steps as to those in \S\ref{sec:se_Fper} are taken to compute $u_{\G}^{f,F}$ and $u_{\Hd}^{f,F}$. However \eqref{eq:se_perint} will no longer contain a periodic wrap of the window function $w$ and it will be defined over the domain $\mathcal{D}$.

When computing the Ewald decomposition for $\G$ (the same discussion holds for $\Hd$), the function $g$ consists of a convolution of the screening function $\gamma_\alpha(\mb{x},\xi)$ as defined in \eqref{eq:screening} and $f_{\G}$,
\begin{align*}
  g(\mb{x}) = \suml_{n=1}^N \gamma_\alpha(\mb{x}-\mb{y}_n,\xi)f_{\G}(\mb{y}_n) = \suml_{n=1}^N \sigma^{n,F}(\mb{x}),
\end{align*}
with $\sigma^{n,F}$ defined in \eqref{eq:sigmadef}.
In this case, the function $g$ is not compactly supported. A truncation level $\epsilon$ and a truncation distance $\delta_1$ is introduced such that $\gamma_\alpha(\mb{x},\xi) < \epsilon$ for $\vert\mb{x}\vert \geq \delta_1/2$.
If the source points, $\mb{y}_n$, are all contained in a domain of size $L\times L$, the computational domain $\D$ must contain the extension $\delta_1$. Moreover, the truncated Gaussians which are used as window functions in the spectral Ewald method, $w(\mb{x},\xi,\eta)$, will also be truncated at $\epsilon$. This introduces another distance $\delta_2$, such that $w(\mb{x},\xi,\eta) < \epsilon$ for $\vert\mb{x}\vert \geq \delta_2$.
The domain $\D$ must therefore extend the original domain of size $L\times L$ such that $\D=\left[-\delta_\epsilon/2, L + \delta_\epsilon/2\right]^2$ for $\delta_\epsilon=\max(\delta_1,\delta_2)$. Defining $\tilde{L} = L + \delta_\epsilon$, the largest point-to-point distance can be computed such that $\R=\sqrt{2}\tilde{L}$.

In order to compute the convolution in \eqref{eq:freespace_phi}, $\G^{\R}$ needs to be defined on a domain of size $2\tilde{L}\times 2\tilde{L}$. Applying an FFT introduces a periodisation of the computations and in order not to pollute the solution within the square $2\tilde{L}\times 2\tilde{L}$ the data needs to be zero-padded. Following the discussion in \cite{AfKlinteberg2017a}, the upsampling factor needed is $s_f \geq 1+\sqrt{2} \approx 2.5$.
Furthermore, to speed up computations and reduce the cost of the extra upsampling needed it is possible to precompute a  mollified Green's function. With $M^2$ grid points as described for the spectral Ewald method in \S\ref{sec:se_Fper} the pre-computation step involves computing the Green's function $\GFR$ on a grid of size $(s_fM)^2$ followed by a 2D IFFT, a truncation in real space to size $(2M)^2$ and finally a 2D FFT back to Fourier space. The remaining computations can then be performed with a plain upsampling factor of two, which is the minimum required for an a-periodic convolution.


\section{Truncation errors}
\label{sec:est}

As the sum in \eqref{eq:sumG_per} has been rewritten into the two rapidly converging sums in \eqref{eq:sumG_per_split} for $\G$ (and with corresponding expressions for $\Hd$), the infinite sums can now be truncated in order to be computationally feasible. The \realS~converges rapidly in physical space, and a cut-off radius $r_c$ is introduced such that the sum is only computed for point pairs within that radius. The \kspaceS~on the other hand converges rapidly in Fourier space and a maximum $k_\infty$ is therefore introduced to limit the sum over $\mb{k}$ to terms $\vert\mb{k}\vert=k\leq k_\infty$. The errors that arise from these truncations are estimated in this section, first for the periodic case and then the free-space case. These estimates will govern the choice of parameters as described in \S\ref{sec:se}.

\subsection{Real space truncation errors}
\label{sec:realest}
The \realS s for the periodic case are defined as
\begin{align}
  u_{\G}^R(\mb{x},\xi) &= \suml_{\mb{p}\in\mathbb{Z}^2}^{*} \suml_{n=1}^N \GR(|\mb{x}-\tau(\mb{p})-\mb{y}_n|,\xi)f_{\G}(\mb{y}_n), \label{eq:Grealsum} \\
  u_{\Hd}^R(\mb{x},\xi) &= \suml_{\mb{p}\in\mathbb{Z}^2}^{*} \suml_{n=1}^N \HdR(\mb{x}-\tau(\mb{p})-\mb{y}_n,\xi)_j f_{\Hd}(\mb{y}_n)_j, \label{eq:Hrealsum}
\end{align}
where $\GR$ and $\HdR$ are defined in \eqref{eq:GR_per} and \eqref{eq:HR_per} respectively. Note that for the free-space case the sum over $\mb{p}$ is removed. The truncation errors of the \realS~arise when limiting the sums above to those $\mb{y}_n$ and $\mb{p}$ such that $|\mb{x}-\tau(\mb{p})-\mb{y}_n|\leq r_c$, for some cut-off radius $r_c$. The real space truncation errors will be the same both in the periodic and free-space cases as the errors are not affected by the removal of the sum over $\mb{p}$.

\subsubsection{$\GR(\mb{r},\xi)$}
\label{sec:realestGR}
\begin{figure}[h!]
  \centering
  \includegraphics[width=0.49\textwidth]{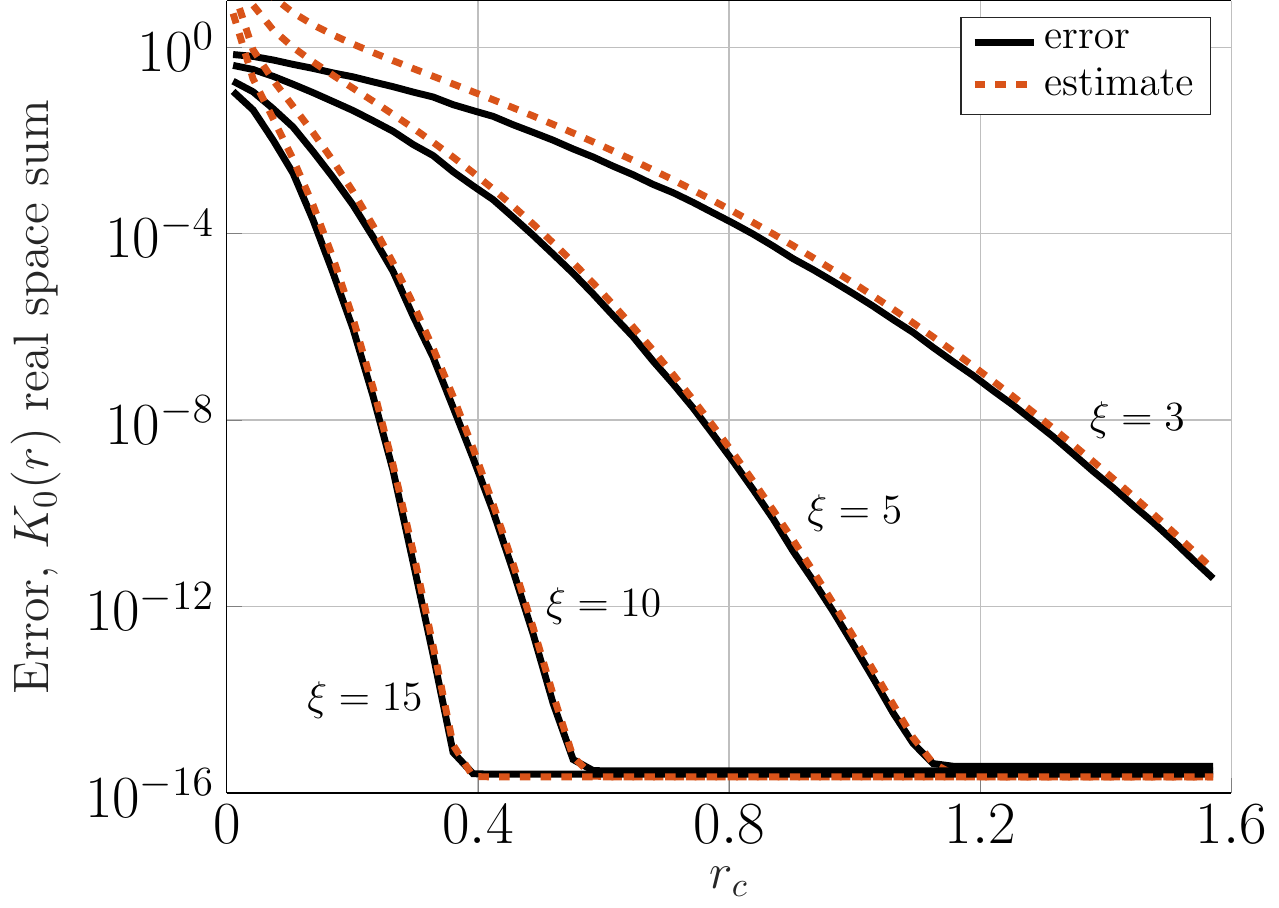}%
  \includegraphics[width=0.49\textwidth]{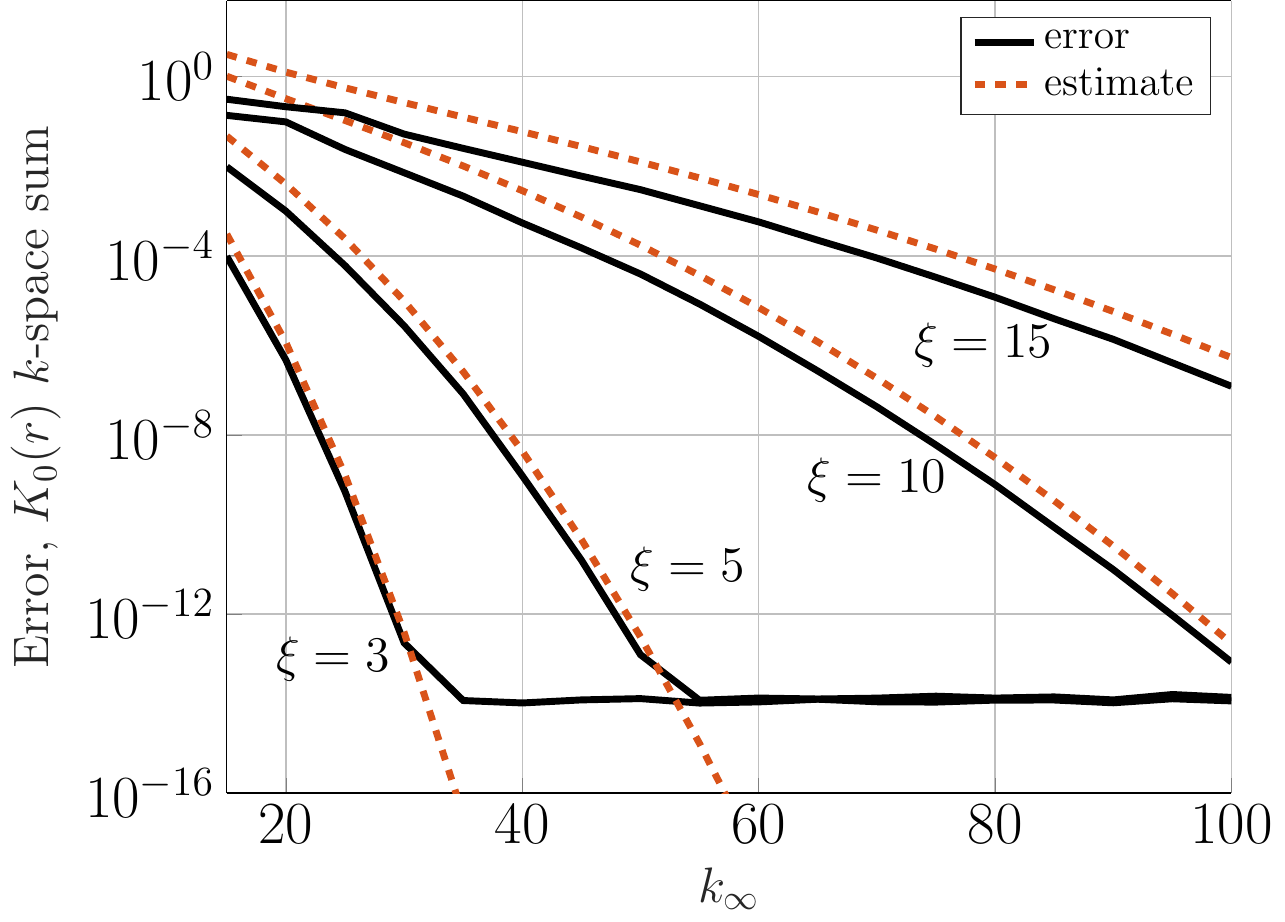}%
  \caption{Truncation errors (absolute) and estimates for different values of $\xi$ for $\G$. The test domain contains $500$ random sources and targets, and random point forces $f_{\G}\in[0,1]$. Left: estimate as derived in \eqref{eq:estRG} for the \realS~for different cut-off radii $r_c$. Right: estimate from \eqref{eq:estkG} for the \kspaceS, when varying $k_\infty$.}
  \label{fig:Gest_2p}
\end{figure}

Following the approach in \cite{AfKlinteberg2017a, Palsson2019} the RMS error, $\delta u_{\G}^R$, can be estimated by
\begin{align*}
  \left( \delta u_{\G}^R\right)^2 \approx \dfrac{1}{L^2} \suml_{n=1}^N f(\mb{y}_n)^2 \il_{r>r_c} \left( \GR(r,\xi)\right)^2d\mb{r},
\end{align*}
where $L$ is the length of one side of the reference cell (periodic case) or the computational domain (free-space case). The domain does not have to be square, but it is assumed here to facilitate notation. Letting $Q_{\G} \coloneqq \sum_{n}f_{\G}(\mb{y}_n)^2$, $\left( \delta u_{\G}^R\right)^2$ can be rewritten as
\begin{align}
  \left( \delta u_{\G}^R\right)^2 \approx \dfrac{Q_{\G}}{L^2} \il_{r>r_c} \left[ \dfrac{1}{2}\Kxy{0}{r^2\xi^2}{\dfrac{\alpha^2}{4\xi^2}} \right]^2 d\mb{r}
  = \dfrac{\pi Q_{\G}}{2L^2} \il_{\rho=r_c}^\infty \Kxy{0}{\rho^2\xi^2}{\dfrac{\alpha^2}{4\xi^2}}^2\rho d\rho,
  \label{eq:duGR1}
\end{align}
where polar coordinates are used and the $\theta$-direction has been integrated. What remains is to approximate the expression in \eqref{eq:duGR1}, for which the definition of $\Kxy{0}{x}{y}$ from \eqref{eq:Kxy_def} is used. Inserting this definition into \eqref{eq:duGR1}, re-arranging the order of the integrals and integrating over $\rho$ gives
\begin{align*}
  \left( \delta u_{\G}^R\right)^2 \approx \dfrac{\pi Q_{\G}}{2L^2} \il_{t=1}^\infty \il_{s=1}^{\infty} \dfrac{e^{-\alpha^2/4\xi^2t}}{t} \dfrac{e^{-\alpha^2/4\xi^2s}}{s} \dfrac{e^{-r_c^2\xi^2(t+s)}}{t+s} ds dt.
\end{align*}
Using the fact that $s,t\geq 1$ and $\alpha^2/4\xi^2 > 0$ the expression can be simplified and approximated as
\begin{align}
  \left( \delta u_{\G}^R\right)^2 \approx \dfrac{\pi Q_{\G}}{4L^2\xi^6r_c^4}e^{-2r_c^2\xi^2}.
  \label{eq:estRG}
\end{align}
The truncation errors and estimates are shown together in Figure~\ref{fig:Gest_2p} (left), for a test domain with $500$ randomly distributed sources and targets, together with random point forces $f_{\G}(\mb{y}_k)\in[0,1]$. In this example, $\alpha=1$ and $L=2\pi$. The estimates hold for varying $L$ and $\alpha$. The estimates follow the error well for error levels below $10^{-2}$. For larger errors, the estimate is not sharp. That is, however, far from the region of interest for an accurate computation.

\subsubsection{$\HdR(\mb{r},\xi)$}
\label{sec:realestHR}
In a similar way as for $\GR$, the truncation-error estimate for $\HdR$ can be estimated by
\begin{align}
  \left(\delta u_{\Hd}^R\right)^2 \approx \dfrac{1}{L^2}\suml_{j=1}^2 \suml_{n=1}^N f_{\Hd}(\mb{y}_n)_j^2 \il_{r>r_c} \left( \HdR(\mb{r},\xi)\right)_j^2d\mb{r}.
  \label{eq:deltauH}
\end{align}
Let $Q_{\Hd}\coloneqq \sum_{j=1}^2 \sum_{n=1}^N f_{\Hd}(\mb{y}_n)_j^2$ and compute
\begin{align*}
  (\overline{\HdR})^2 = \dfrac{1}{2}\suml_{j=1}^2 \left(\HdR\right)_j^2 = \dfrac{\xi^4}{2\alpha^2}r^2\Kxy{-1}{r^2\xi^2}{\dfrac{\alpha^2}{4\xi^2}}^2.
\end{align*}
Inserted into \eqref{eq:deltauH} this gives
\begin{align*}
  \left(\delta u_{\Hd}^R\right)^2 \approx  \dfrac{Q_{\Hd}\xi^2}{2L^2\alpha^2}\il_{r>r_c} r^2 \Kxy{-1}{r^2\xi^2}{\dfrac{\alpha^2}{4\xi^2}}^2 d\mb{r} = \dfrac{Q_{\Hd}\xi^2 \pi}{L^2\alpha^2} \il_{\rho=r_c}^\infty \rho^3 \Kxy{-1}{\rho^2\xi^2}{\dfrac{\alpha^2}{4\xi^2}}^2d\rho,
\end{align*}
again using polar coordinates and integrating in the $\theta$-direction. As for $\GR$, using the definition of $\Kxy{-1}{x}{y} = \int_{1}^\infty e^{-xt-y/t}dt$ and re-arranging the integration order gives, after integrating over $\rho$,
\begin{align*}
   \left(\delta u_{\Hd}^R\right)^2 \approx \dfrac{Q_{\Hd}\pi}{2L^2\alpha^2} \il_{t=1}^\infty \il_{s=1}^\infty \dfrac{e^{-\alpha^2/4\xi^2(t+s)} e^{-r_c^2\xi^2(t+s)}(1+r_c^2\xi^2(t+s))}{(t+s)^2}ds dt.
\end{align*}
Again, using that $s,t\geq 1$ and $\alpha^2/4\xi^2 > 0$ it is obtained
\begin{align}
\left( \delta u_{\Hd}^R\right)^2 \approx \dfrac{\pi Q_{\Hd}}{2L^2\alpha^2}\dfrac{e^{-2r_c^2\xi^2}(3+2r_c^2\xi^2)}{r_c^4\xi^4} \approx \dfrac{\pi Q_{\Hd}}{L^2\alpha^2 r_c^2\xi^2}e^{-2r_c^2\xi^2},
\label{eq:estRH}
\end{align}
where in the last step only leading order terms of $r_c\xi$ have been kept.

\begin{figure}[h!]
  \centering
  \includegraphics[width=0.49\textwidth]{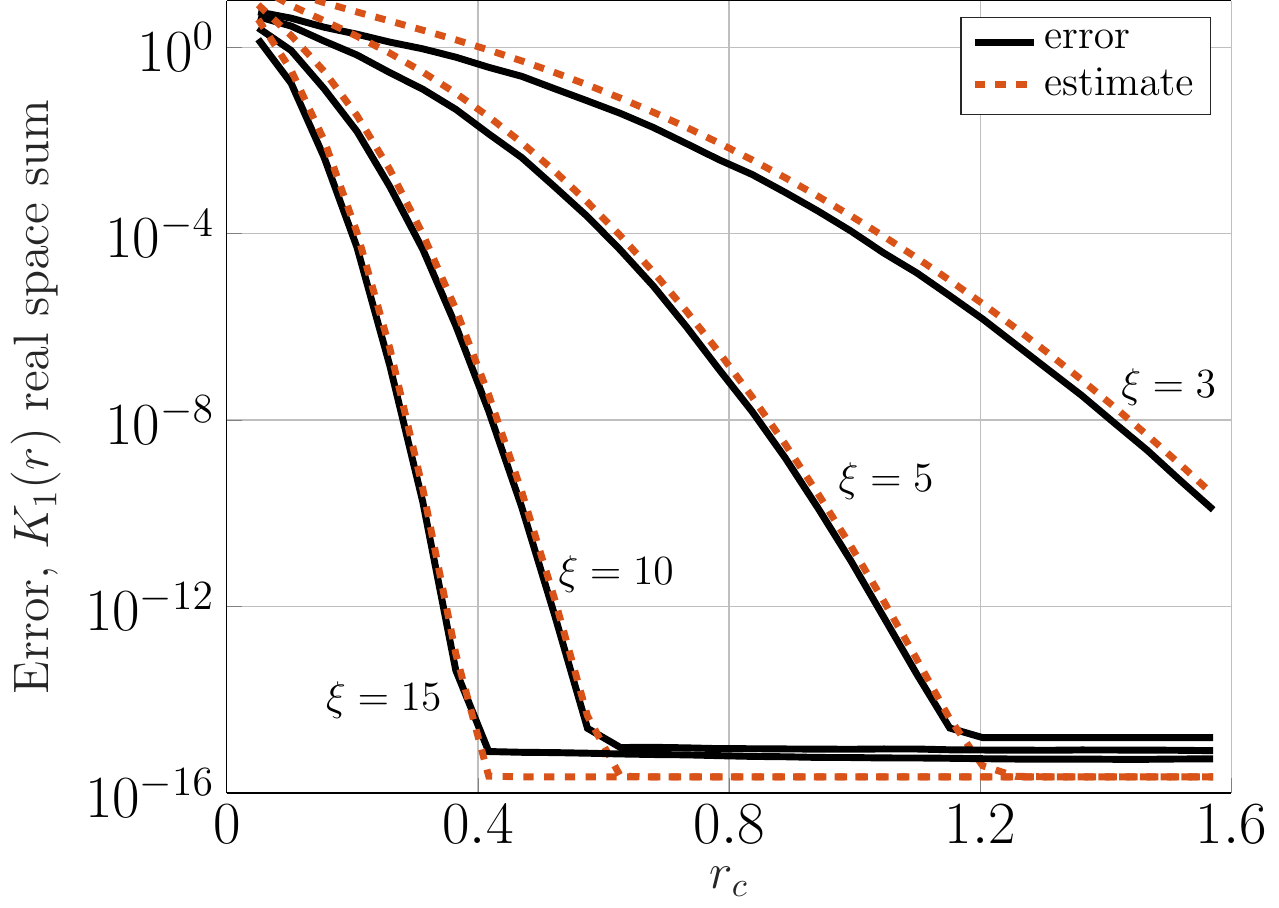}%
  \includegraphics[width=0.49\textwidth]{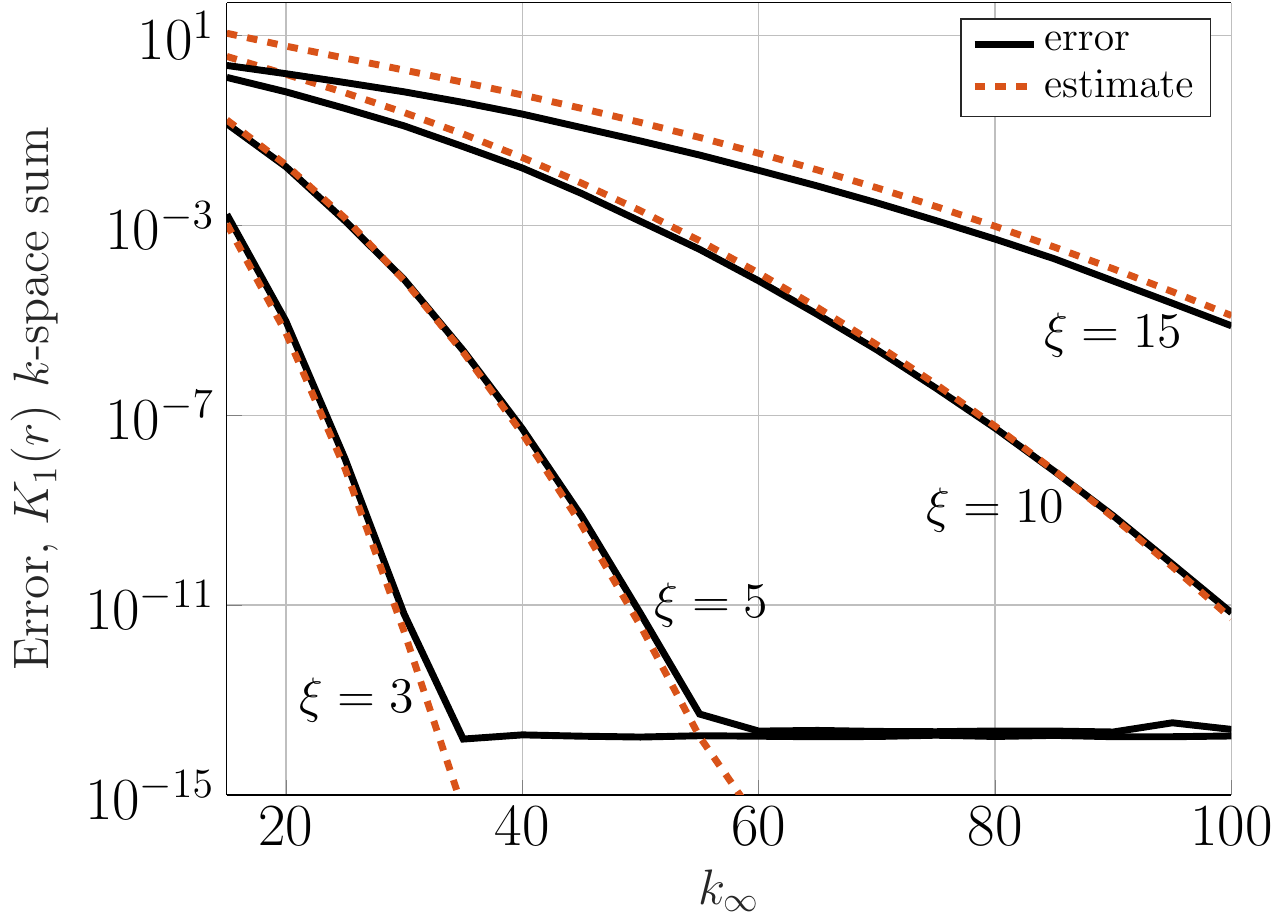}%
  \caption{Truncation errors (absolute) and estimates for different values of $\xi$ for $\Hd$. The test domain contains $500$ random sources and targets, and random point forces $f_{\G}\in[0,1]$. Left: estimate as derived in \eqref{eq:estRH} for the \realS~for different cut-off radii $r_c$. Right: estimate from \eqref{eq:estkH} for the \kspaceS, when varying $k_\infty$.}
  \label{fig:Hest_2p}
\end{figure}

In Figure~\ref{fig:Hest_2p} (left) the truncation error and estimate are shown for different values of $\xi$. The test domain has $500$ randomly distributed sources and targets and random point sources $\mb{f}_{\Hd}(\mb{y}_k)\in[0,1]$. For this example, $\alpha=1$ and $L=2\pi$ but the estimate holds also for varying $\alpha$ and $L$. Similar to the estimates for $\GR$, the estimate follow the error well for error levels below $10^{-2}$.

\subsection{Fourier space truncation errors}
The \kspaceS s are in the periodic case defined as
\begin{align}
  u_{\G}^F(\mb{x},\xi) &= \dfrac{1}{L^2}\suml_{\mb{k}\neq 0} \GF(\mb{k},\xi)\suml_{n=1}^N f_{\G}(\mb{y}_n)e^{-i\mb{k}\cdot(\mb{x}-\mb{y}_n)}, \label{eq:Gksum} \\
  u_{\Hd}^F(\mb{x},\xi) &= \dfrac{1}{L^2}\suml_{\mb{k}\neq 0} \HdF(\mb{k},\xi)_j \suml_{n=1}^N f_{\Hd}(\mb{y}_n)_j e^{-i\mb{k}\cdot(\mb{x}-\mb{y}_n)}. \label{eq:Hksum}
\end{align}
Truncation errors arise when the sums over $\mb{k}$ are truncated at some $k_\infty$ such that $k \leq k_\infty$ for $k=\vert\mb{k}\vert$. The estimates are derived first for the periodic case and then altered to better follow the errors in the free-space case.

Following the work in \cite{Kolafa1992}, let
\begin{align*}
  E(\mb{x}) = \suml_{n=1}^N q_n \left(f(\mb{x}-\mb{x}_n) - \tilde{f}(\mb{x}-\mb{x}_n)\right),
\end{align*}
be an error measure due to a set of pointwise errors. It then holds that the RMS error, $\delta E$, can be approximated as
\begin{align}
  \delta E^2 \approx \dfrac{1}{|\tilde{V}|}\suml_{n=1}^N q_n^2 \il_{\tilde{V}}\left(f(\mb{r})-\tilde{f}(\mb{r})\right)^2 d\mb{r},
\end{align}
where $\tilde{V}$ is a disc enclosing all point-to-point vectors $\mb{r}_{jl} = \mb{x}_j-\mb{x}_l$.

\subsubsection{$\GF(\mb{k},\xi)$}
\label{sec:GFest}
The error from truncating the discrete Fourier sum at $k_\infty$ has the form
\begin{align*}
  u_{\G}^F - \tilde{u}_{\G}^F  = \dfrac{1}{L^2} \suml_{n=1}^N f_{\G}(\mb{y}_n) \suml_{\mb{k}, k>k_\infty}
  \GF(\mb{k},\xi)  e^{i\mb{k}\cdot(\mb{x}-\mb{y}_n)}
\end{align*}
The RMS error, $\delta u_{\G}^F$, is estimated as
\begin{align*}
  \left(\delta u_{\G}^F\right)^2 \approx \suml_{n=1}^N f(\mb{y}_n)^2 \dfrac{1}{|\tilde{V}|} \il_{\tilde{V}}|e_{\G}(\mb{r})|^2 d\mb{r},
\end{align*}
where $\tilde{V}$ is a disc with radius $L/2$. The pointwise error is approximated as follows
\begin{align*}
  e_{\G}(\mb{r}) \coloneqq \dfrac{1}{L^2} \suml_{\mb{k}, k>k_\infty} \GF(\mb{k},\xi) e^{i\mb{k}\cdot\mb{r}} \approx  \dfrac{1}{L^2} \il_{k>k_\infty} \GF(\mb{k},\xi)e^{i\mb{k}\cdot\mb{r}}d\mb{k},
\end{align*}
where the sums over $k_1,k_2$ are approximated by a double integral \cite{Kolafa1992}.
Inserting the expression for $\GF$ from \eqref{eq:GF_per}, switching to polar coordinates
\begin{align*}
    \begin{cases}
      \mb{k} = \kappa\left(\cos(\theta), \sin(\theta)\right), \\
      \mb{r} = r\left(\cos(\beta),\sin(\beta)\right),
    \end{cases}
\end{align*}
and integrating over $\theta$, gives
\begin{align*}
  \vert e_{\G}(\mb{r})\vert \approx \dfrac{2\pi}{L^2}\il_{k>k_\infty} \dfrac{e^{-(\alpha^2+k^2)/4\xi^2}}{\alpha^2 +k^2}e^{i\mb{k}\cdot\mb{r}}d\mb{k} =
   \dfrac{4\pi^2}{L^2}\il_{\kappa=k_\infty}^\infty  \dfrac{\kappa e^{-(\alpha^2+\kappa^2)/4\xi^2}}{\alpha^2+\kappa^2}\Jx{0}{\kappa r} d\kappa.
\end{align*}
The remaining integral is difficult to compute. In order to compute it, first the trick of \cite{Tornberg2016} is again used, letting $\lambda \coloneqq 1/4\xi^2$ and differentiating $e_{\G}(\mb{r})$ with respect to $\lambda$:
\begin{align*}
  \dfrac{\partial \vert e_{\G}\vert}{\partial \lambda} \approx -\dfrac{4\pi^2}{L^2}\il_{\kappa=k_\infty}^\infty \kappa e^{-(\alpha^2+\kappa^2)\lambda}\Jx{0}{\kappa r}d\kappa.
\end{align*}
To estimate the integral, $\Jx{0}{x}$ is approximated as $\Jx{0}{x} \sim \dfrac{\sqrt{2}}{\sqrt{\pi x}}$ for large values of $x$, which gives
\begin{align*}
  \dfrac{\partial \vert e_{\G}\vert}{\partial \lambda} \approx  -\left(\dfrac{\pi^3 2^5}{r L^4} \right)^{1/2} \il_{\kappa = k_\infty}^\infty \sqrt{\kappa}e^{-(\alpha^2+\kappa^2)/4\xi^2}d\kappa  =
   -\left(\dfrac{\pi^3 2^5}{r L^4} \right)^{1/2}\dfrac{e^{-\alpha^2\lambda}}{2\lambda^{3/4}}\Gamma\left(\dfrac{3}{4},k_\infty^2 \lambda\right),
\end{align*}
where $\Gamma(s,x)$ is the incomplete Gamma function. For large $x$ it can be estimated as $\Gamma(3/4,x)\sim e^{-x}/x^{1/4}$. Thus, $\frac{\partial \vert e_{\G}\vert}{\partial \lambda}$ is simplified to
\begin{align*}
  \dfrac{\partial \vert e_{\G}\vert}{\partial \lambda} \approx  -\left(\dfrac{\pi^3 2^5}{r L^4} \right)^{1/2} \dfrac{e^{-(\alpha^2+k_\infty^2)\lambda}}{2\lambda\sqrt{k_\infty}}.
\end{align*}
To obtain an approximation of $\vert e(\mb{r})\vert$, integrate w.r.t. $\lambda$,
\begin{align*}
  \vert e_{\G}(\mb{r})\vert \approx& \il_0^{\lambda} -\left(\dfrac{\pi^3 2^5}{r L^4} \right)^{1/2} \dfrac{e^{-(\alpha^2+k_\infty^2)\rho}}{2\rho\sqrt{k_\infty}}d\rho =
  \il_{\lambda}^\infty \left(\dfrac{\pi^3 2^5}{r L^4} \right)^{1/2} \dfrac{e^{-(\alpha^2+k_\infty^2)\rho}}{2\rho\sqrt{k_\infty}}d\rho  \\
  &=
  \left(\dfrac{\pi^3 2^5}{r L^4} \right)^{1/2} \dfrac{\Gamma\left(0,(\alpha^2+k_\infty^2)\lambda\right)}{\sqrt{k_\infty}},
\end{align*}
where the integration limits can be switched as the limit of the integrand is zero when $\lambda$ approaches infinity. Expanding $\Gamma(0,x)$ for large $x$ gives $\Gamma(0,x)\sim e^{-x}/x$, and the expression above can thus be simplified to
\begin{align*}
  \vert e_{\G}(\mb{r})\vert \approx \left(\dfrac{\pi^3 2^3}{r L^4 k_\infty \lambda^2} \right)^{1/2} \dfrac{e^{-(\alpha^2+k_\infty^2)\lambda}}{(\alpha^2+k_\infty^2)}.
\end{align*}
Using $Q_{\G}$ from \S\ref{sec:realestGR} the truncation-error estimate becomes
\begin{align}
  \left(\delta u_{\G}^F\right)^2 \approx \dfrac{512 Q_{\G} \pi^3 \xi^4}{L^5(\alpha^2+k_\infty^2)^2k_\infty}e^{-2(\alpha^2+k_\infty^2)/4\xi^2}.
  \label{eq:estkG}
\end{align}
Using the same example as in  \S\ref{sec:realestGR}, the estimate and truncation errors are plotted together in Figure~\ref{fig:Gest_2p} (right). Note here that although the estimates do not scale as well with $\xi$ for the \kspace~as for the \rspace~estimates, they always over estimate the errors.

\subsubsection{$\HdF(\mb{k},\xi)$}
The approach to estimate the truncation for $\Hd$ follows the same pattern. First, the truncation error for the \kspaceS~can be expressed as
\begin{align*}
  u_{\Hd}^F - \tilde{u}_{\Hd}^F = \dfrac{1}{L^2}\suml_{n=1}^N f_{\Hd}(\mb{y}_n)_j \suml_{\mb{k}, k>k_\infty} \HdF(\mb{k},\xi)_j e^{i\mb{k}\cdot(\mb{x}-\mb{y}_n)},
\end{align*}
and the RMS error approximated by
\begin{align}
  (\delta u_{\Hd}^F)^2 \approx \sum_{n=1}^N f_{\Hd}(\mb{y}_n)^2_j \dfrac{1}{|\tilde{V}|}\il_{\tilde{V}}|e_{\Hd}(\mb{r})_j|^2d\mb{r},
  \label{eq:RMSestHFdef}
\end{align}
using the same notation as in \S\ref{sec:GFest}. Define
\begin{align*}
  \vert e_{\Hd}(\mb{r})_j\vert  \coloneqq \dfrac{1}{L^2}\suml_{\mb{k},k>k_\infty} \HdF(\mb{k},\xi)_j e^{i\mb{k}\mb{r}}
  \approx \dfrac{1}{L^2}\il_{k>k_\infty} \HdF(\mb{k},\xi)_j e^{i\mb{k}\mb{r}}d\mb{k}.
\end{align*}
Inserting the expression for $\HdF$ as defined in \eqref{eq:HF_per}, gives
\begin{align}
  \vert e_{\Hd}(\mb{r})_j\vert  \approx \dfrac{2\pi i}{\alpha L^2} \il_{k>k_\infty} \dfrac{\hat{k}_j k}{\alpha^2 + k^2}e^{-(\alpha^2+k^2)/4\xi^2}e^{i\mb{k}\cdot\mb{r}}d\mb{k},
  \label{eq:eHd1}
\end{align}
where $\hat{k}_j \coloneqq k_j/k$. To estimate $e_{\Hd}$, first approximate
\begin{align*}
  \hat{k}_j \approx \sqrt{\dfrac{1}{2}\suml_{j=1}^2 \hat{k}_j^2} = \dfrac{1}{\sqrt{2}}.
\end{align*}
Inserting this expression into \eqref{eq:eHd1}, switching to polar coordinates and integrating in the $\theta$-direction yields
\begin{align*}
  \vert e_{\Hd}(\mb{r})_j\vert  \approx \dfrac{4\pi^2 i}{\alpha L^2 \sqrt{2}}\il_{\kappa = k_\infty}^\infty \dfrac{\kappa^2}{\alpha^2+\kappa^2}e^{-(\alpha^2+\kappa^2)\lambda} J_0(\kappa r)d\kappa,
\end{align*}
where $\lambda \coloneqq 1/4\xi^2$ similarly as in \S\ref{sec:GFest}. Again, differentiating with respect to $\lambda$  and inserting the approximation $\Jx{0}{x}\sim \sqrt{\frac{2}{\pi x}}$ gives the expression
\begin{align*}
  \dfrac{\partial \vert e_{\Hd,j}\vert}{\partial \lambda} \approx -\dfrac{4\pi^2 i}{\alpha L^2 \sqrt{\pi r}} \il_{\kappa = k_\infty}^{\infty} \kappa^{3/2}e^{-(\alpha^2+\kappa^2)\lambda} d\kappa
  = -\dfrac{4\pi^2 i}{\alpha L^2 \sqrt{\pi r}} \dfrac{e^{-\alpha^2 \lambda}}{2\lambda^{5/4}}\Gamma\left( \dfrac{5}{4}, k_\infty^2\lambda\right).
\end{align*}
Using the approximation $\Gamma(\frac{5}{4},x) \sim e^{-x}x^{1/4}$ for large $x$, it can be approximated as
\begin{align*}
    \dfrac{\partial \vert e_{\Hd,j}\vert }{\partial \lambda} \approx -\dfrac{2\pi^2 i}{\alpha L^2 \sqrt{\pi r}}\dfrac{\sqrt{k_\infty}e^{-(\alpha^2+k_\infty^2)\lambda}}{\lambda}
    \Rightarrow e_{\Hd}(\mb{r})_j \approx \dfrac{2\pi^2 i}{\alpha L^2 \sqrt{\pi r}} \sqrt{k_\infty} \Gamma\left(0,(\alpha^2+k_\infty^2)\lambda\right),
\end{align*}
when integrating with respect to $\lambda$. Again, using $\Gamma(0,x)\sim e^{-x}/x$ gives
\begin{align*}
  |e_{\Hd}(\mb{r})_j| \approx \dfrac{2\pi^2}{\alpha L^2 \sqrt{\pi r}} \dfrac{\sqrt{k_\infty}e^{-(\alpha^2+k_\infty^2)\lambda}}{(\alpha^2+k_\infty^2)\lambda}.
\end{align*}
The RMS error can then be approximated by \eqref{eq:RMSestHFdef} as
\begin{align}
   (\delta u_{\Hd}^F)^2 \approx \dfrac{128\pi Q_{\Hd} k_\infty \xi^4}{L^5 \alpha^2 (\alpha^2+k_\infty^2)^2}e^{-2(\alpha^2+k_\infty^2)/4\xi^2},
   \label{eq:estkH}
\end{align}
by integrating over the disc $\tilde{V}$ and using $Q_{\Hd}$ as defined in \S\ref{sec:realestHR}. The errors and estimates for $\xi=3,5,10,15$ are shown in Figure~\ref{fig:Hest_2p} (right) for the example described in \S\ref{sec:realestHR}.

\subsection{Truncation-error estimates for the free-space case}
For the free-space case, the discrete Fourier sum is replaced by a Fourier transform in $u_{\G}^F$ and $u_{\Hd}^F$ as is shown in \eqref{eq:sumG_fs_split}. The real-space truncation error remains unchanged and thus the estimate for the real-space truncation error is unchanged. For the \kspaceP, it is noted that the behaviour of the additional terms in $u_{\Hd}^F$ depends on the parameters of the problem $\alpha$ and $\R$,
\begin{align*}
  \begin{cases}
  \uf_{\G}(\mb{x}) - \tilde{u}^f_{\G}(\mb{x}) &= \dfrac{1}{(2\pi)^2}\il_{k>k_\infty} \GFR(\mb{k},\xi)\suml_{n=1}^N f_{\G}(\mb{y}_n)e^{i\mb{k}\cdot(\mb{x}-\mb{y}_n)}, \\
  \uf_{\Hd}(\mb{x}) - \tilde{u}^f_{\Hd}(\mb{x}) &= \dfrac{1}{(2\pi)^2}\il_{k>k_\infty} \HdFR(\mb{k},\xi)_j\suml_{n=1}^N f_{\Hd}(\mb{y}_n)_je^{i\mb{k}\cdot(\mb{x}-\mb{y}_n)}.
\end{cases}
\end{align*}
For the applications motivating this paper, $\R=\mathcal{O}(10)$ and $\alpha\gg1$.  For these values of $\alpha$ and $\R$ the estimates remain unchanged, as the extra terms in $\GFR$ are negligible. The only scaling to consider is then the difference in scaling of the Fourier transform. For other ranges of parameters $\R$ and $\alpha$ however, free-space specific estimates are needed. Their derivation follows that of the periodic case.

For $\GFR$, the following expression needs to be estimated,
\begin{align*}
  e_{\G}^{\R}(\mb{r}) \coloneqq \dfrac{1}{(2\pi)^2} \il_{k>k_\infty} \GFR(\mb{k},\xi) e^{i\mb{k}\cdot\mb{r}}.
\end{align*}
Using the definition of $\GFR$ in \eqref{eq:GF_fs}, a switch to polar coordinates $(\kappa,\theta)$ and integration in the $\theta$-direction gives the following expression of $e_{\G}^{\R}$
\begin{align*}
  \vert e_{\G}^{\R}(\mb{r})\vert  = \il_{\kappa = k_\infty}^\infty
  \left[ \dfrac{1 + \alpha\kappa \Jx{1}{\kappa \R}\Kx{0}{\alpha\R}-\alpha\R\Jx{0}{\kappa\R}\Kx{1}{\alpha\R}}{\alpha^2+\kappa^2}\right]
  \kappa \Jx{0}{\kappa r}e^{-(\alpha^2+\kappa^2)/4\xi^2}d\kappa.
\end{align*}
This can be divided into three expressions $e_1$, $e_2$ and $e_3$, each computed with the same techniques as for the periodic case. Using the following approximations:
$\Jx{0}{x}\sim \sqrt{\frac{2}{\pi x}}$, $\Jx{1}{x}\sim - \sqrt{\frac{2}{\pi x}}$ and $\Gamma\left(0,x\right) \sim e^{-x}/x$ for large $x$ the RMS of the truncation error can be approximated as
\begin{align*}
  \left(\delta u_{\G}^{\R,F}\right)^2 &\approx \dfrac{64 Q_{\G} \xi^4}{L(\alpha^2+k_\infty^2)^2}e^{-2(\alpha^2+k_\infty^2)/4\xi^2}\left(\dfrac{1}{\sqrt{2\pi k_\infty}} - \dfrac{\alpha \Kx{0}{\alpha\R}}{\sqrt{\R}\pi} - \dfrac{\alpha\sqrt{\R}\Kx{1}{\alpha\R}}{\pi k_\infty} \right)^2.
\end{align*}

Similarly, for $\HdFR$, the expression to estimate reads
\begin{align*}
  e_{\Hd}^{\R}(\mb{r}) \coloneqq \dfrac{1}{(2\pi)^2} \il_{k>k_\infty} \HdFR(\mb{k},\xi) e^{i\mb{k}\cdot\mb{r}}.
\end{align*}
Again, using the definition of $\HdFR$ in \eqref{eq:HF_fs} and following the steps in the periodic case, one arrives at the expression
\begin{align*}
  \left(\delta u_{\Hd}^{\R,F}\right)^2 &\approx \dfrac{8Q_{\Hd}\xi^2}{L\pi^2\alpha^2}
  \dfrac{e^{-2(\alpha^2+k_\infty^2)/4\xi^2}}{(\alpha^2+k_\infty^2)^2}
  \left( \sqrt{2\pi k_\infty} -\dfrac{8\alpha K_0(\alpha\R)k_\infty}{\sqrt{\R}} -2\alpha\sqrt{\R}K_1(\alpha\R)\right)^2.
\end{align*}


\section{Numerical results}
\label{sec:res}
This section collects results regarding the Ewald decomposition and its computation. First the complexity of the method is demonstrated and second the evaluation on a uniform grid is shown. Lastly, the dependence of the Ewald decomposition of $\Kx{0}{\alpha r}$ and $\Kx{1}{\alpha r}$ with respect to the parameter $\alpha$ is discussed.

The \realP s~(both $\GR$ and $\HdR$) are computed efficiently using algorithms by \citeauthor{harris2009} \cite{harris2009}. These algorithms are correct up to an absolute accuracy of $10^{-10}$.

\subsection{Computational complexity}
\begin{figure}[h!]
  \centering
  \includegraphics[width=0.49\textwidth]{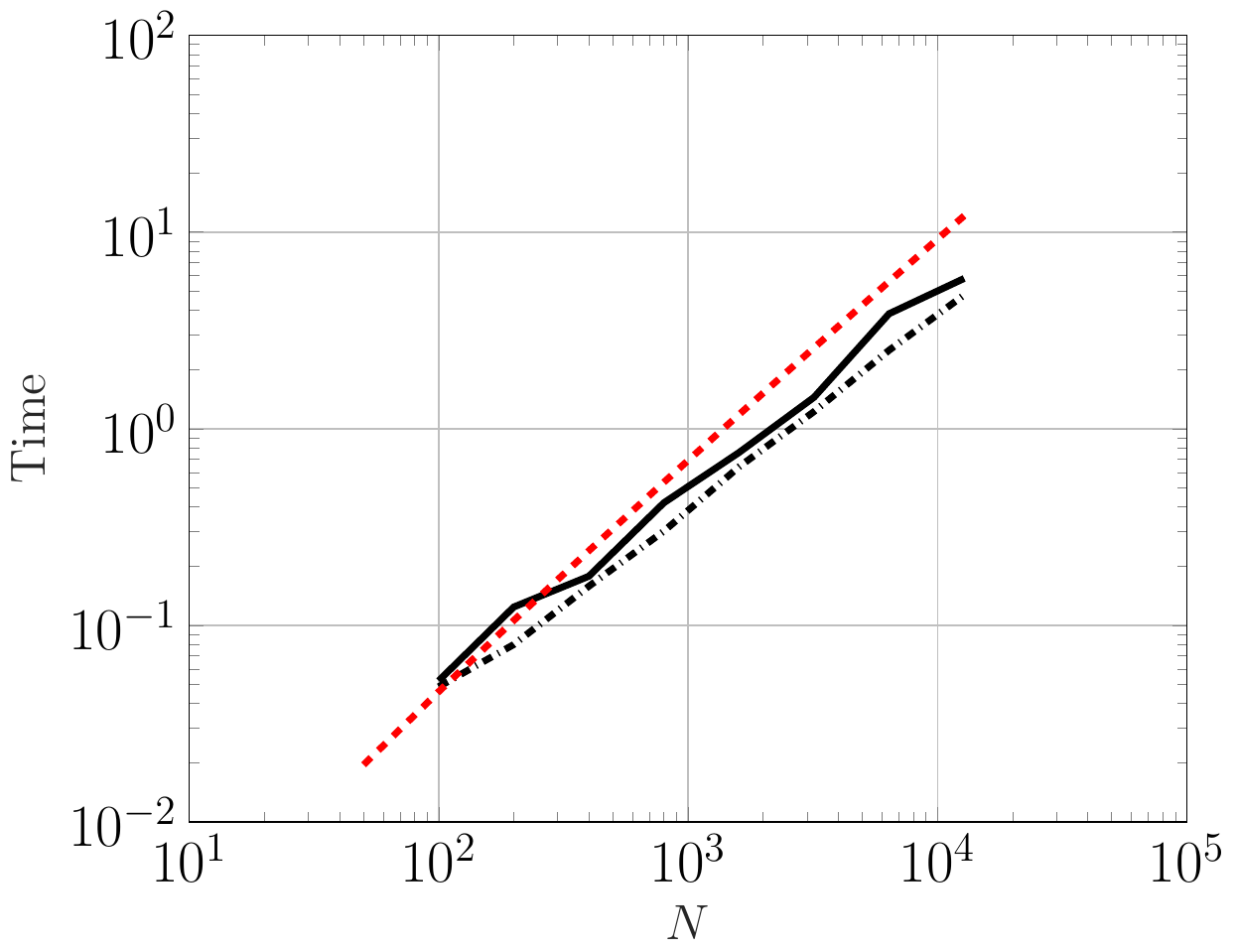}
  \includegraphics[width=0.49\textwidth]{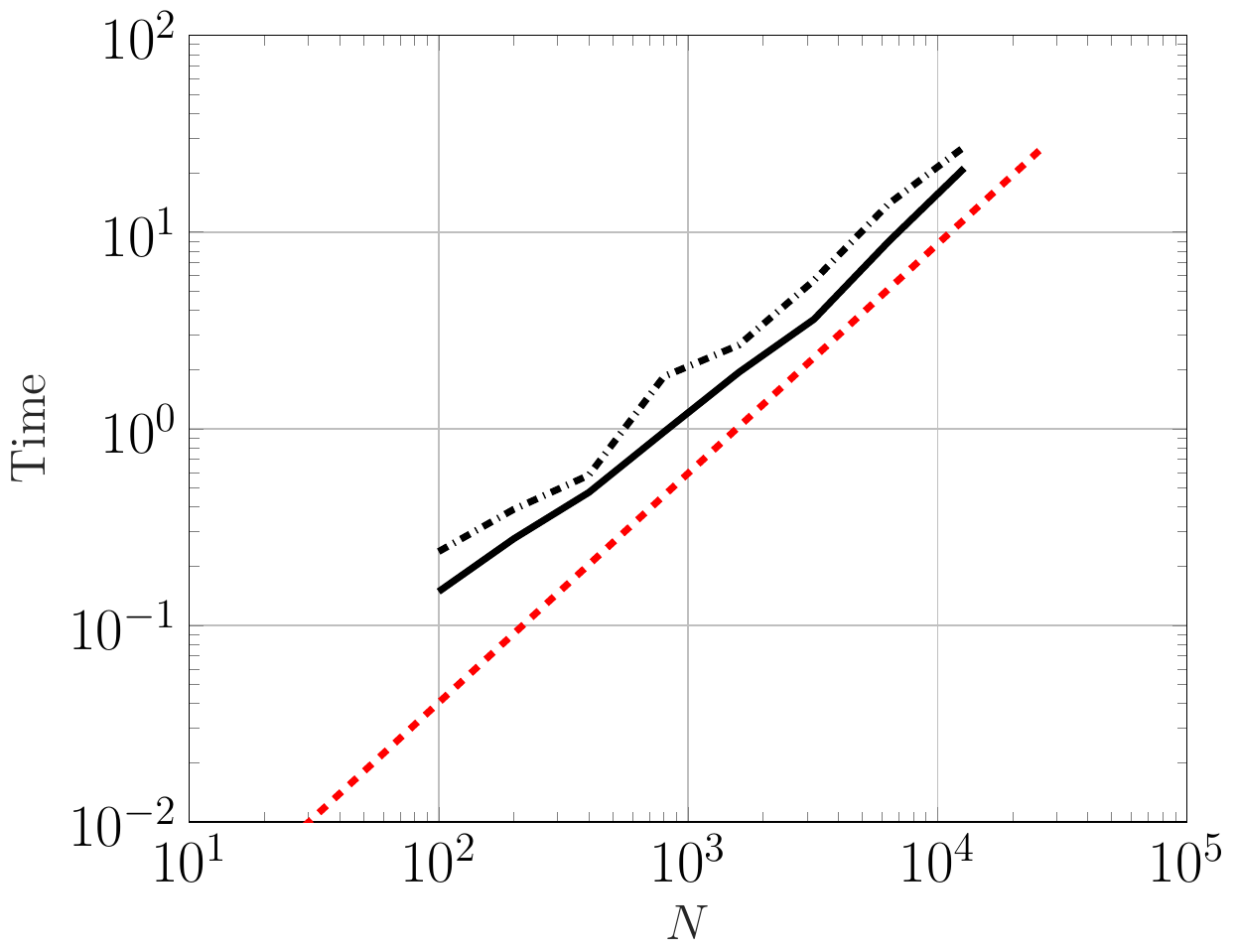}
  \caption{Complexity for the spectral Ewald method when computing $u_{\G}$ (solid black lines) and $u_{\Hd}$ (dot-dashed black lines) for $N$ random sources and targets. Left: periodic case. Right: free-space case. Chosen parameters for this simulation are $L=2\pi$, $\alpha=1$ and $p=24$. The red, dashed line is a reference line for $\mathcal{O}(N\log N)$ complexity.}
  \label{fig:se_timings}
\end{figure}

To demonstrate the computational complexity of the spectral Ewald method, a MATLAB implementation of the method is used to compute $u_{\G}$ and $u_{\Hd}$ for a varying number $N$ of random targets in both the free-space and periodic case. In this example the number of sources equals the number of targets. Using the estimates in \S\ref{sec:est}, $r_c$, $\xi$ and $k_\infty$ are chosen to keep a constant number of neighbours within the \rspace~cut-off radius. Other parameters are $L=2\pi$, $\alpha=1$ and $p=24$. How the method scales can be seen in Figure~\ref{fig:se_timings} for the periodic case (left) and the free-space case (right), where it is shown to be a little faster than $\mathcal{O}(N\log N)$.
The implementation which is used here is not optimised, therefore the scaling and not the constant is of interest.

\subsection{On grid evaluation}
In certain applications, the discrete sums of \eqref{eq:sumG} and \eqref{eq:sumH} are evaluated for targets on a uniform grid, for example in \cite{fryklund2019}.

When all targets $\mb{x}_t$ are on a uniform grid with grid spacing $\tilde{h}$, it is possible to take advantage of the uniform grid already introduced in the \emph{spreading step} of the spectral Ewald method in \S\ref{sec:se}. The grid on which $H(\mb{x})$ from \eqref{eq:se_H} is evaluated is chosen such that $h=L/M$ is an integer multiple of $\tilde{h}$, or vice versa. For such a grid, the \emph{quadrature step} in \S\ref{sec:se} where information is gathered from the grid to the target points is superfluous.
The scheme changes accordingly: the sum $u^{F,P}$ in \eqref{eq:se_persum} is first rewritten as
\begin{align*}
    u^{F,P}(\mb{x}_t) = \dfrac{1}{L^2}\suml_{\mb{k}} \widehat{A}^F(\mb{k},\xi)e^{-i\mb{k}\cdot\mb{x}_t}\dfrac{1}{\widehat{w}_k}\suml_{n=1}^N f(\mb{y}_n)\widehat{w}_ke^{i\mb{k}\cdot\mb{y}_n}, \; t=1,\hdots N.
\end{align*}
In comparison with the expression in \eqref{eq:se_ufp_w} for the case of non-uniform targets, this expression contains only one power of the window function $\widehat{w}_k$. Using the same definition of $H$ as in \eqref{eq:se_H}, the scaling in \eqref{eq:se_Htildek} is modified into
\begin{align*}
  \widehat{\tilde{H}}(\mb{k}) = \widehat{A}^F(-\mb{k},\xi)\dfrac{\widehat{H}(\mb{k})}{\widehat{w}_k}.
\end{align*}
The last step in computing $u^{F,P}$ can be seen as computing
\begin{align*}
  u^{F,P} = \dfrac{1}{L^2}\suml_{\mb{k}} \widehat{\tilde{H}}(-\mb{k})e^{-i\mb{k}\cdot\mb{x}_t},
\end{align*}
for all target points $\mb{x}_t$, which can be accomplished efficiently using a 2D inverse FFT. The modification to on-grid evaluation is here stated for the periodic case, but follows the same pattern in the free-space case.

\begin{figure}[h!]
  \centering
  \includegraphics[width=0.4\textwidth]{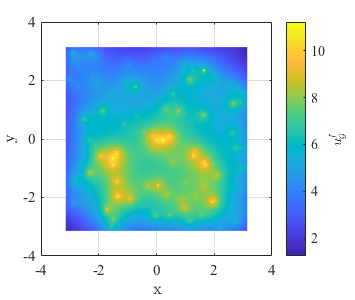}
  \includegraphics[width=0.4\textwidth]{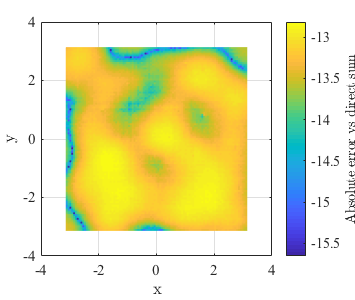}
  \caption{Left: $u_{\G}^{f}$ evaluated at $100\times 100$ uniform target points for $100$ randomly placed sources, with random point strengths $f_{\G}\in[0,1]$. Right: $\log_{10}$ of the absolute error of the solution obtained by the spectral Ewald method compared to a direct sum.}
  \label{fig:se_grid}
\end{figure}

The removal of the last quadrature step speeds up the computations and maintains the same error levels as a straightforward application of the spectral Ewald method.
In Figure~\ref{fig:se_grid}, the solution of $u^{f}_{\G}$ is evaluated at $100\times 100$ uniform target points and $100$ sources for random source point strengths $f_{\G}$ (left) together with the error of the spectral Ewald method compared to the direct sum. The modified approach for grid evaluation reduces the computational time by $15\%$. In this example, $\xi$, $r_c$ and $k_\infty$ are set to an error tolerance of $10^{-12}$ using the truncation-error estimates. Other parameters are $L=2\pi$, $\alpha=1$ and $p=24$.

\subsection{Modified Greens function for the free-space case}
\label{sec:results_modG}
In the free-space case, the task of computing $\uf_{\G}$ and $\uf_{\Hd}$ from \eqref{eq:sumG} and \eqref{eq:sumH} respectively faces different challenges depending on the value of the parameter $\alpha$ in \eqref{eq:modhelm}.
 To demonstrate, an example of evaluating $\uf_\G$ with $N_s=100$ source and target points with random source forces $f_{\G}\in[0,1]$ is considered. The functions $\G(r)=\Kx{0}{\alpha r}$ and $\Hd(r)=\Kx{1}{\alpha r}\mb{r}/r$ depend on two variables: the difference between a source and target point, $\mb{r}$, and the parameter $\alpha$.

To compute the $\mb{k}$-space part in the free-space case corresponds to computing the integral in \eqref{eq:sumG_fs_split} rather than a discrete sum as for the periodic case. This integral contains the term $1/(\alpha^2+k^2)$. Consequently, care needs to be taken when $\alpha$ is small as the integral becomes nearly singular around $k=0$. The modified Green's function $\GFR(k,\xi)$ removes the near-singularity by introducing a finite limit for $k=0$ when $\alpha$ approaches $0$.

\begin{figure}[h!]
  \centering
  \includegraphics[width=0.49\textwidth]{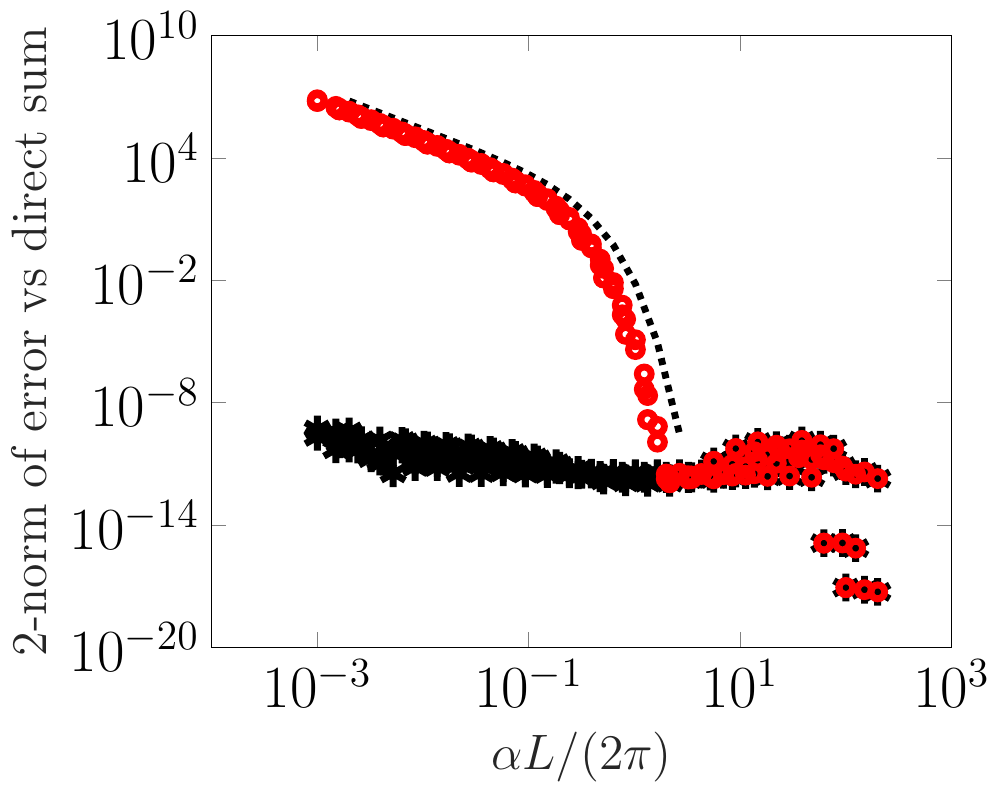}
  \includegraphics[width=0.49\textwidth]{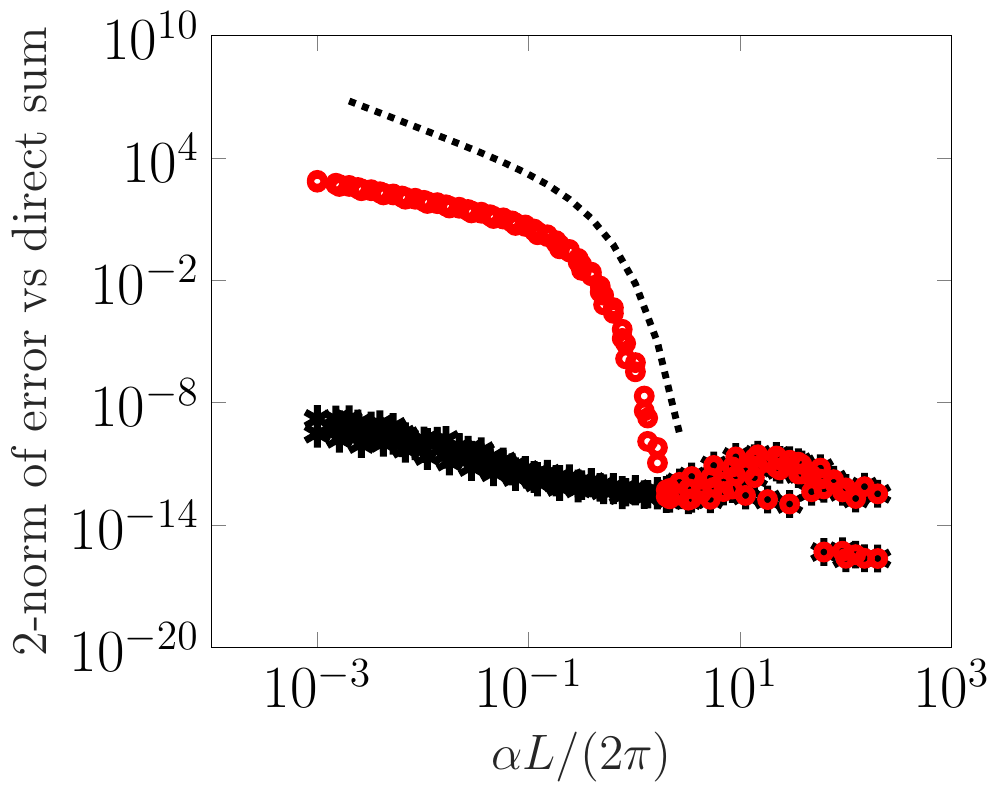}
  \caption{Error of solution computed with the spectral Ewald method compared to the direct sum using the original $\GF$ (red circles) and the modified $\GFR$ (black asterisks) for $\G$ and $\Hd$ respectively (left and right). The dashed black line corresponds to $\GF(0,\xi)-\GFR(0,\xi)$ in both figures.}
  \label{fig:se_alpha}
\end{figure}

As an example, in Figure~\ref{fig:se_alpha} the absolute error compared to a direct sum of computing $\Kx{0}{\alpha r}$ (left) and $\Kx{1}{\alpha r}$ (right) with the spectral Ewald method using $\GF$ and $\HdF$
(red circles) is compared to using $\GFR$ and $\HdFR$ (black asterisks). The parameters for the example has been varied as  $L=2\pi$, $3\pi$ and $4\pi$ together with $\xi=5$ and $10$. It is clear that the point where $\GF$ needs to be changed to $\GFR$ (and $\HdF$ to $\HdFR$) occurs for similar $\alpha L/2\pi$, independently of $\xi$.
This point can be approximated by regarding the difference $\GF(0,\xi)-\GFR(0,\xi)$, which in Figure~\ref{fig:se_alpha} is the black, dashed line for $\xi=10$, $L=4\pi$ (other values of $\xi$ and $L$ yield similar lines). For $\Hd$ this over estimates the error slightly, but the term $\HdF(0,\xi)-\HdFR(0,\xi)=0$ always and cannot be used. Through numerical tests, the point where the modified Green's functions are needed is estimated as $\alpha L/2\pi \lesssim 1.5$.

For values of $\alpha L/2\pi >1.5$, the difference between $\GF$ and $\GFR$ (and also between $\HdF$ and $\HdFR$) is small. Therefore, it is sufficient to use the standard $\GF(k,\xi)$. As the periodic case in \eqref{eq:sumG_per_split} consists of evaluating discrete sums over $\mb{k}$ rather than integrals, the issue of a near singularity for small $\alpha$ never arises. Therefore, no special treatment is needed in the periodic case.

\begin{figure}[h!]
  \centering
  \includegraphics[width=0.4\textwidth]{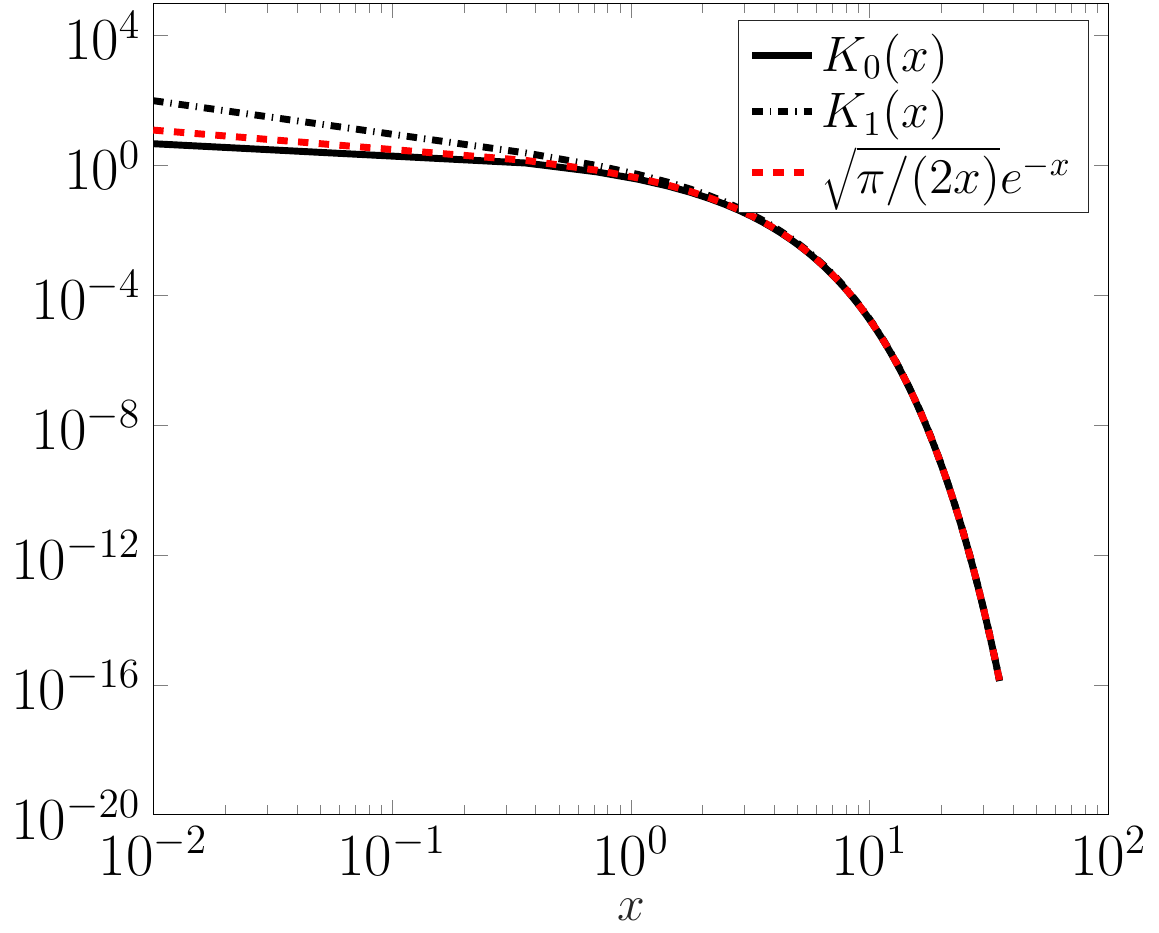}
  \caption{Decay of $\Kx{0}{x}$ (solid black line) and $\Kx{1}{x}$ (dashed black line). The red line corresponds to $\sqrt{\pi/(2x)}e^{-x}$.}
  \label{fig:se_K0K1decay}
\end{figure}

Furthermore, it is worth noting that both $\Kx{0}{x}\rightarrow0$ and $\Kx{1}{x}\rightarrow 0$ when $x\rightarrow \infty$. For large arguments, they both tend to $\Kx{\nu}{x} \sim \sqrt{\pi/(2x)}e^{-x}$ for $\nu=0,1$, as is shown in Figure~\ref{fig:se_K0K1decay}.
Thus, for values of $\alpha$ sufficiently large the sums \eqref{eq:sumG} and \eqref{eq:sumH} will converge rapidly.
It may then be beneficial to consider a cut-off radius for the original sums, rather than using the Ewald decomposition. Such a cut-off radius, $\tilde{r}$, can be obtained by solving $\sqrt{\pi/(2\alpha \tilde{r})}e^{-\alpha\tilde{r}}>\epsilon$ for some truncation level $\epsilon$. The truncated sums can then be computed using a similar approach as that of the \realS, with a neighbour-list implementation. Contrary to the case of the spectral Ewald method, where work can be shifted to the $\mb{k}$-space sum in order to keep the \realS~$\mathcal{O}(N)$ cost, such an implementation would remain $\mathcal{O}(N^2)$. With the quickly decaying Green's function, however, the constant would be small.

\section{Conclusions}
In this paper, Ewald decompositions of $\Kx{0}{\alpha r}$ and $\Kx{1}{\alpha r}$ in two dimensions have been derived both for periodic and non-periodic problems. The spectral Ewald method was used to compute solutions efficiently. To further decrease the computational cost, special treatment for on-grid evaluation was considered. Moreover, the dependence of the parameter $\alpha$ on the computations was discussed. Truncation-error estimates were derived that approximate the errors that arise when truncating both the real-space and $\mb{k}$-space sums. These estimates were used to compute optimal parameters needed for the spectral Ewald method.

\section{Acknowledgements}
This work is supported by the G{\"o}ran Gustafsson Foundation for Research in Nature and Medicine. A.K.T. also gratefully acknowledges the support from the Swedish Research Council, Grant no. 2015-04998.

\newpage

\bibliographystyle{plainnat}
\bibliography{../Thesis/saraslib_abb.bib}

\end{document}